# The Antisymmetry of Distortions

Brian K. VanLeeuwen and Venkatraman Gopalan

Materials Research Institute and Department of Materials Science and Engineering, Pennsylvania State University, University Park, PA, 16802

**Distortions are ubiquitous in nature. Under perturbations such as stresses, fields, or other changes, a physical system reconfigures by following a path from one *state* to another; this path, often a collection of atomic trajectories, describes a *distortion*. Here we introduce an antisymmetry operation called *distortion reversal*, 1\*, that reverses a distortion pathway. The symmetry of a distortion pathway is then uniquely defined by a distortion group involving 1\*; it has the same form as a magnetic group that involves time reversal, 1'. Given its isomorphism to magnetic groups, distortion groups could potentially have commensurate impact in the study of distortions as the magnetic groups have had in the study of magnetic structures. Distortion symmetry has important implications for a range of phenomena such as structural and electronic phase transitions, diffusion, molecular conformational changes, vibrations, reaction pathways, and interface dynamics.**

A *distortion* or *distortion pathway* refers to any one of the many possible paths between two or more states of a system. They are important for understanding chemical reaction kinetics[1–5], phonon modes[6–8], molecular pseudorotations and conformational changes[3–5], diffusion[9–12], the motion of interfaces such as grain boundaries[13,14], domain walls[15,16], and dislocations[17,18], and emergent phenomena in transient or metastable states that arise from a distortion of the ground state[19–24]. There is often a privileged point on a pathway that is extremal in energy. For stable phonons, this is the *ground state*; for unstable phonons, the *parent structure*; for reaction pathways, the *transition state*; and for phase transitions, the *saddle point*. The relative energy of this point corresponds to the activation energy in transition state theory[3,25]. In many important pathways, it is found that the energies on either side of this privileged point are symmetric, *e.g.* when opposite sides are mirror images of each other. An antisymmetry operation named *distortion reversal*, 1\* is introduced here to describe the complete symmetry of such pathways (the coloring of operations coupled with 1', 1\*, and 1'* is intended to assist the reader and has no special meaning beyond that). When combined with conventional symmetry groups, it gives rise to *distortion groups*. The symmetry of a distortion pathway is uniquely described by a distortion group.

Distortions, especially phonon modes, are studied today using representation analysis[6–8], through decomposition onto a symmetry-adapted basis using irreducible representations (irreps). Why then is the concept of distortion-reversal symmetry and distortion groups necessary? It is instructive to look at the history, where a similar question has been posed for over 45 years regarding the need for time-reversal symmetry and magnetic groups versus representation analysis for the study of magnetic structures[26,27]. Opechowski & Dreyfus rigorously showed that representation analysis of magnetic structures and magnetic groups were equivalent, through a correspondence between 1-dimensional real irreps and magnetic groups[26,28,29]. In practice however, magnetic groups are widely used today due to their ease of use in visualizing complex spin structures, easy transformations for predicting the form of magnetic property tensors and in deriving the energy invariants in magnetic crystals in a simple and transparent manner. In contrast, distortions and vibrations in molecules and crystals are studied today *only* by representation analysis. There is currently no equivalent formalism to the time-reversal symmetry or the magnetic groups for studying distortions. This work provides that framework through the introduction of distortion-reversal symmetry and distortion groups.

In developing distortion symmetry, we discovered that a somewhat similar concept was introduced several decades ago in transition state theory in the limited context of reversing reactants and products in simple molecular reactions[3–5]. We demonstrate here that the concept of distortion groups is much more general, and can be used for studying "distortions" interpreted in a very broad sense, including in crystals, interfaces, electronic structure and diffusive systems. Further, they can predict the form of tensors that describe any property change of a system as a function of distortion.

We introduce the concept of distortion reversal in Fig. 1a in a discrete system through three randomly placed atoms (in red) that form the parent structure. The atoms then displace to their new positions as per the displacements shown as arrows. The final distorted structure (in light pink) is the result of displacing each position accordingly. The action of the distortion-reversal operation, 1*, on the distortion in Fig. 1a is the reversal of displacements $u_i$ of the atoms $i$ (=1, 2, 3) to $-u_i$. These displacements have been decomposed in Fig. 1c into rotation ($u_{i,R}$, Fig. 1d), translation ($u_{i,T}$, Fig. 1e), scaling ($u_{i,S}$, Fig. 1f), and deformation ($u_{i,D}$, Fig. 1g), i.e. $u_i = u_{i,R} + u_{i,T} + u_{i,S} + u_{i,D}$. This is analogous to the Helmholtz decomposition of continuous vector fields into components (see Methods). This decomposition is not necessary for implementing 1*, but it is helpful to illustrate the relationship between 1*, and the rotation-reversal operation, $1^\Phi$, introduced by Gopalan and Litvin[30]. While $1^\Phi$ reverses the rotation component, $u_{i,R}$ (Fig. 1e) to $-u_{i,R}$, it has no clear implications for the other components. This creates a problem in implementing $1^\Phi$, because it requires the identification of appropriate polyhedra within a structure that exhibit rigid rotations, but not the other components; the process for such polyhedral identification is non-unique, and often approximate in real systems. In this work, no such polyhedron is required to be identified in implementing 1* as seen from Figs. 1a and 1b. Further, 1* *reverses all the components of* $u_i$, i.e. $1^*(u_{i,R}, u_{i,T}, u_{i,S}, u_{i,D}) = (-u_{i,R}, -u_{i,T}, -u_{i,S}, -u_{i,D})$, not just rotation, $u_{i,R}$, and in this sense, $1^\Phi$ is a special case of 1*. There is an alternate way to view the action of 1* as described below, which will be used in the rest of this article. For linear atomic paths of atoms indexed by subscript $i$, the final atomic positions ($r'_i$) are given by $r'_i = r_i + \lambda u_i$, where $-1 \leq \lambda \leq +1$ and $r_i$ are the initial positions of atoms, $i$, in the intermediate state (Fig. 1h). A typical distortion pathway may begin at $\lambda = -1$, go through an intermediate state at $\lambda = 0$, and end at $\lambda = +1$. We can reverse this pathway by reversing the parameter $\lambda \rightarrow -\lambda$ while leaving the displacement amplitude $u_i$ constant. The atomic trajectories in this example are linear with respect to $\lambda$, but in general, the pathway can be a nonlinear function $r'_i(\lambda)$; the action of 1* will then be to reverse this function $r'_i(\lambda) \rightarrow r'_i(-\lambda)$.

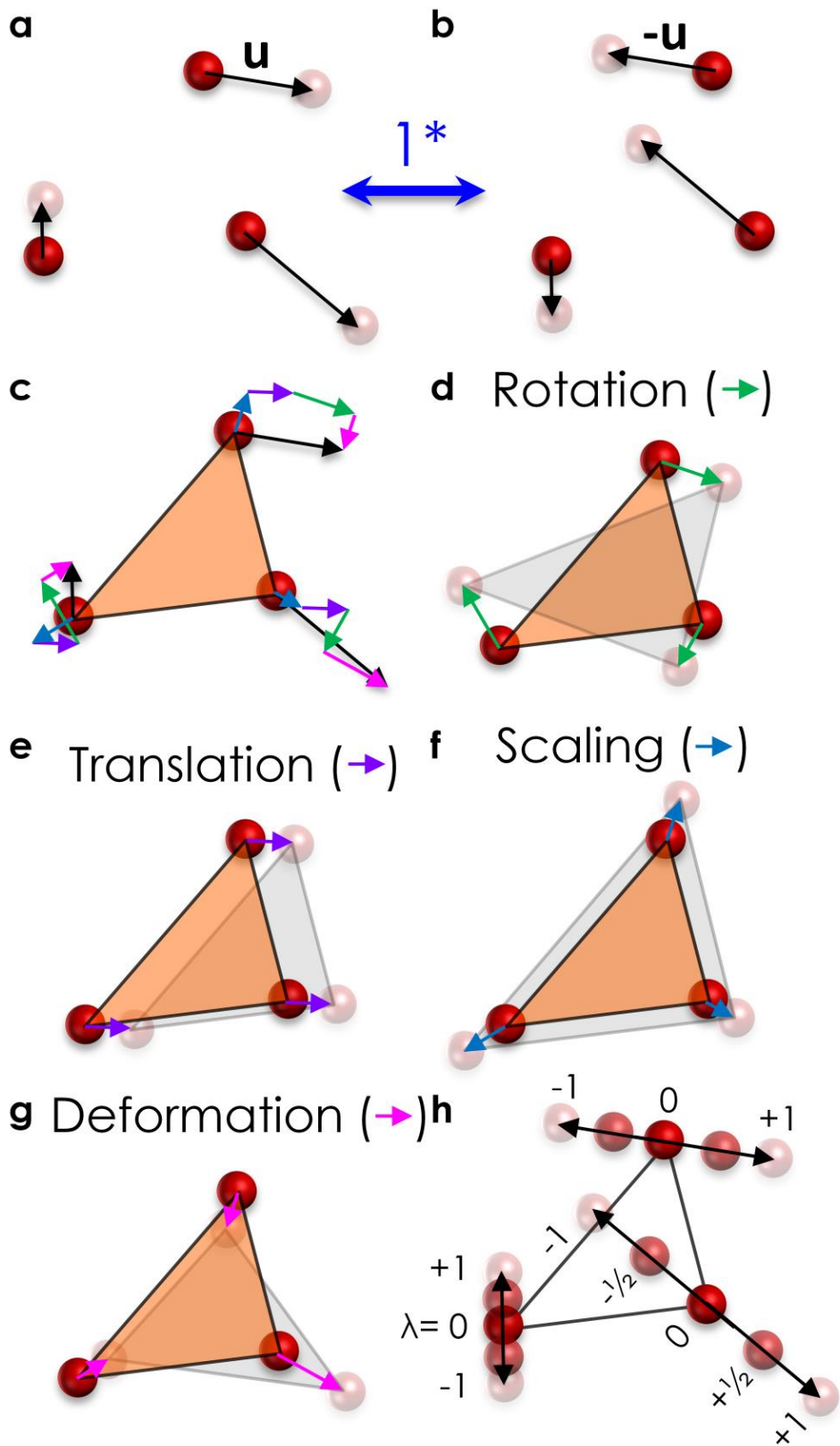

**Figure 1: A simple example of a distortion and its decomposition.** Three atoms are displaced to new positions in **a**. 1* reverses these displacements in **b**. In **c**, the displacements are decomposed into rotation (**d**), translation (**e**), scaling (**f**), and deformation (**g**). The linear trajectories are parameterized by $\lambda$ in **h**.

We now make several important observations regarding the distortion-reversal operation, $1^*$. First, in addition to the ordinary dimensions, a distortion has a *time-like* dimension, $\lambda$, that describes the extent of the distortion. For a reaction pathway, this is the reaction coordinate; for a phonon mode, this is the amplitude; and for a phase transition, this is the order parameter. Note that for coordinates $(\boldsymbol{r}, t, \lambda)$, spatial inversion, $\bar{1}$, reverses the position vector $\boldsymbol{r} \rightarrow -\boldsymbol{r}$, $1'$ reverses the time $t \rightarrow -t$, and $1^*$ reverses $\lambda \rightarrow -\lambda$. Secondly, an analogy to the time-reversal operation, $1'$ can be drawn from Fig. 1. If $\lambda$ is replaced with $t$, then the displacement vectors $\boldsymbol{u}$ are replaced with velocity vectors, $\boldsymbol{v}$, and $1^*$ is replaced by $1'$ between panels in Fig. 1a and 1b. If the velocities were decomposed in a similar way as in Fig. 1, the rotational component of this decomposition would correspond to the angular momentum, and for charged particles, magnetic moment. Because it was inspired by the practice of applying $1'$ to reverse the localized magnetic moments of atoms, Gopalan and Litvin's rotation-reversal operation, $1^\Phi$, focused exclusively on the rotational component[30]. Third, we note that the action of $1^*$ is well defined on any structure that is parameterized by $\lambda$, not just a system of discrete atomic positions and displacements. For example, in calculating ferroelectric polarization, the modern theory of polarization implicitly parameterizes the electronic structure of a system by $\lambda$ by parameterizing the ionic positions, and then calculates the ground state electronic structure for a series of steps between $0 \leq \lambda \leq +1$ [31]. On such a system, $1^*$ has a well-defined action, even on the electronic structure itself (see later discussion on Berry phase and Supplementary Discussion 5). Fourth, we note that the symmetry of a distortion is *not* the symmetry of any particular static structure along the pathway, but rather of the symmetry of the entire pathway. The distortion group maps the entire pathway onto itself, and *not* the individual structures onto themselves; an infinitesimal section of the pathway may map to another section of the pathway through a distortion group such that the pathway as a whole remains invariant.

We first demonstrate distortion symmetry in a distortion of a simple molecule and show how it can predict relevant property changes. Figure 2 shows the pseudorotation distortion of phosphorus pentafluoride, $PF_5$, a well-known fluxional molecule. The ground state geometry of $PF_5$ has $\bar{6}2m$ symmetry. The distortion proceeds by the Berry mechanism[4] where the pair of fluorine atoms on the high symmetry axis move down as another pair of fluorine atoms move up. The structure goes through an intermediate transition state with 4mm symmetry to a final state with $\bar{6}2m$ symmetry. Although this distortion is not a rotation, the final state is equivalent to the original structure rotated by 90°, hence the term "pseudorotation". We calculated the minimum energy pathway (MEP) using the nudged elastic band (NEB) method[32]. The MEP represents the set of most likely trajectories that atoms will follow when physically transitioning between these states, and NEB calculations discretize the distortion pathway into a sequence of "images". The highest energy point on the MEP is known as the transition state and corresponds to $\lambda = 0$ in Fig. 2. The energy of the transition state corresponds to the activation energy.

The MEP of the PF$_5$ pseudorotation was determined to have 4*mm* symmetry (see Methods section). A polynomial fit to the energy profile, $\Delta E$, of the MEP, shown in Fig. 2a, is symmetric, i.e. it is invariant under $\lambda \to -\lambda$. This is required by the distortion group, 4*mm*, as shown next. Because 1* commutes with all spatial operations, the action of any starred symmetry operation on a coefficient of the power series expansion of $P(\lambda)$ can be determined, where $P$ is any property of the system. Specifically, the energy, $P = \Delta E$ in Fig. 2 is a scalar property and is invariant under rotation. By applying tensor transformation rules, we find that 4*$\Delta E(\lambda)$ =$\Delta E(-\lambda)$. However, since 4* is a symmetry operation, Neumann's principle[33,34] states that 4*$\Delta E(\lambda)$ =$\Delta E(\lambda)$. Equating the two, one obtains

$$\Delta E(\lambda) = \Delta E(-\lambda). \qquad (1)$$

In other words, $\Delta E(\lambda)$ is a symmetric function of $\lambda$, which is consistent with the fit in Fig. 2. The three bond lengths, PF1, PF2, and PF3, also follow the requirements of the 4*mm* symmetry as shown in Fig. 2. For example, 4*(PF1($\lambda$))=PF1($-\lambda$), 4*(PF2($\lambda$))=PF3($-\lambda$), and 4*(PF3($\lambda$))=PF2($-\lambda$). Since 4* is a symmetry of the distortion, by Neumann's principle, PF1($\lambda$)=PF1($-\lambda$), PF2($\lambda$)=PF3($-\lambda$), and PF3($\lambda$))=PF2($-\lambda$). This is consistent with the results of the NEB calculation shown in Fig. 2b. Supplementary Discussion 1 presents a similar application of distortion groups to vibrations in H$_2$O and NH$_3$ molecules.

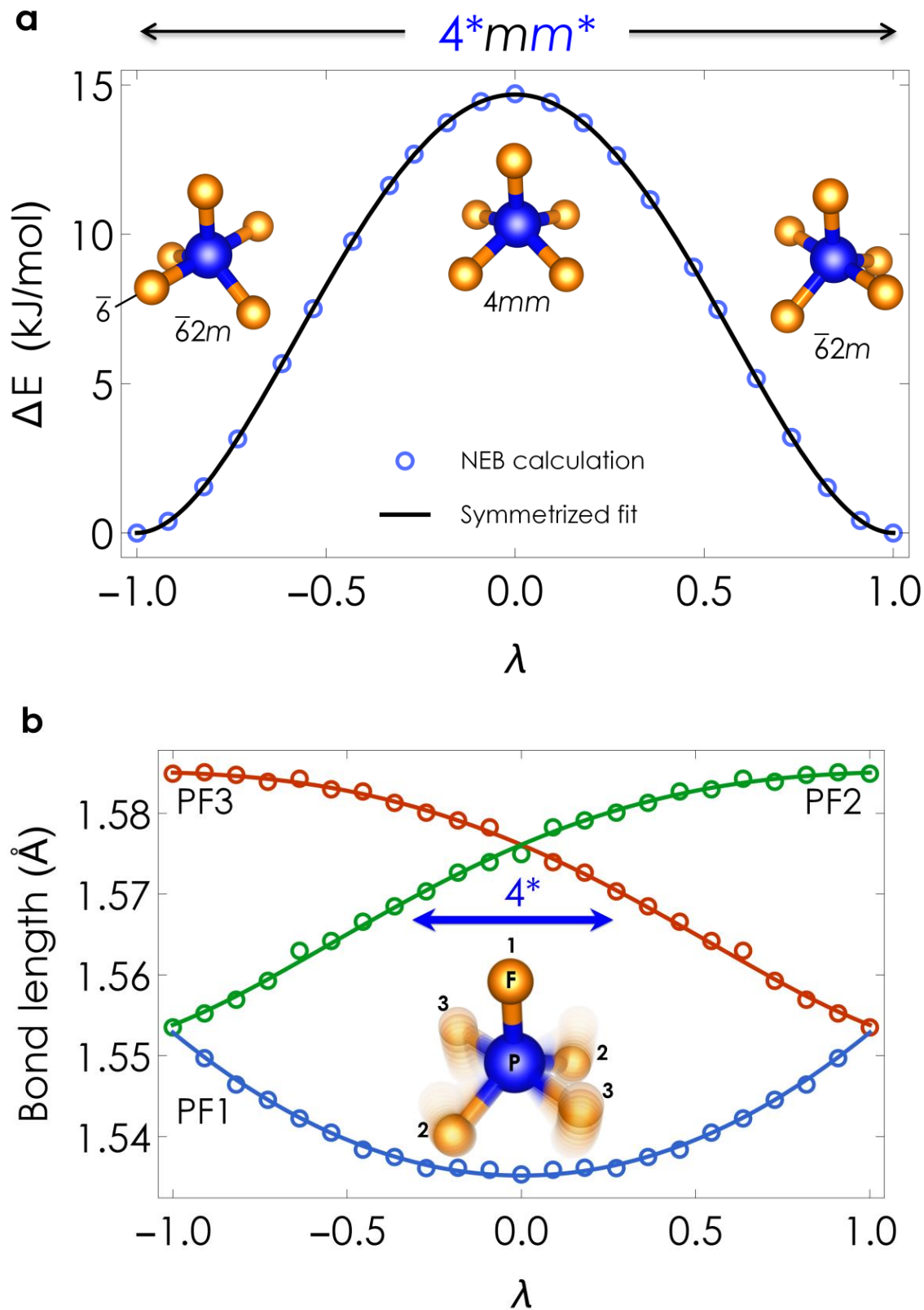

**Figure 2: Symmetry of the PF₅ pseudorotation.** The symmetric energy profile in **a** is guaranteed by the 4*mm* symmetry of the minimum energy pathway (MEP) of the PF₅ pseudorotation. In following this MEP from $\lambda = -1$ to $+1$, any static structure at some fixed value of $\lambda$ can be transformed into the static structure at $-\lambda$ by a 4-fold rotation along the PF1 bond, thus this MEP is invariant under 4*. The PF1 bond length function plotted in **b** is also guaranteed to be symmetric and PF2 and PF3 are required to be mirror images by the 4*mm* symmetry.

Next we demonstrate a symmetry-based approach to testing the stability of a pathway and checking the results of numerical computations for accuracy. This is demonstrated in the NEB calculation of activation energy for an oxygen atom diffusing across a $C_6$ ring on the surface of graphene (Fig. 3). Linear interpolation from the state with oxygen on the right ($\lambda=-1$), to the state with oxygen on the left ($\lambda=+1$) creates a path with *m*m2** symmetry, with a high activation energy barrier; this is not a minimum energy pathway. Linear interpolation like this is the typical method of creating an initial path for NEB calculations. Relaxing this path using NEB cannot and does not change the *m*m2** symmetry, because every NEB iteration must conserve distortion symmetry (Supplementary Discussion 2). We can now systematically explore perturbations to this path by using the irreps of *m*m2** summarized by the following character table:

|  | 1 | 2* | m | m* | Kernel |
|---|---|---|---|---|---|
| $\Gamma_1$ | 1 | 1 | 1 | 1 | *m*m2** |
| $\Gamma_2$ | 1 | 1 | -1 | -1 | 2* |
| $\Gamma_3$ | 1 | -1 | 1 | -1 | m |
| $\Gamma_4$ | 1 | -1 | -1 | 1 | m* |

Similar to the methods of mode crystallography, we can use these irreps to construct a symmetry-adapted basis (Supplementary Discussion 1) from an arbitrary basis set for general perturbations of the path. Using perturbations associated with the irreps, $\Gamma_2$, $\Gamma_3$, and $\Gamma_4$, we can reduce the symmetry of our initial guess path to the symmetry of their kernels, 2*, m, and m* respectively. To achieve a trivial symmetry (point group 1) path, we can combine these. For this example in Figure 3, subspaces associated with $\Gamma_2$ and $\Gamma_3$ are "stable", while for the subspace associated with $\Gamma_4$, one or more directions are "unstable", i.e. small perturbations of the path in these directions will decrease the energy of the path and there will be a net force driving the path away from *m*m2**. This is similar to the unstable phonons of an unstable structure. The stability and instability of any path could clearly be calculated using a method analogous to finite displacement methods used for calculating phonon frequencies of static structures.

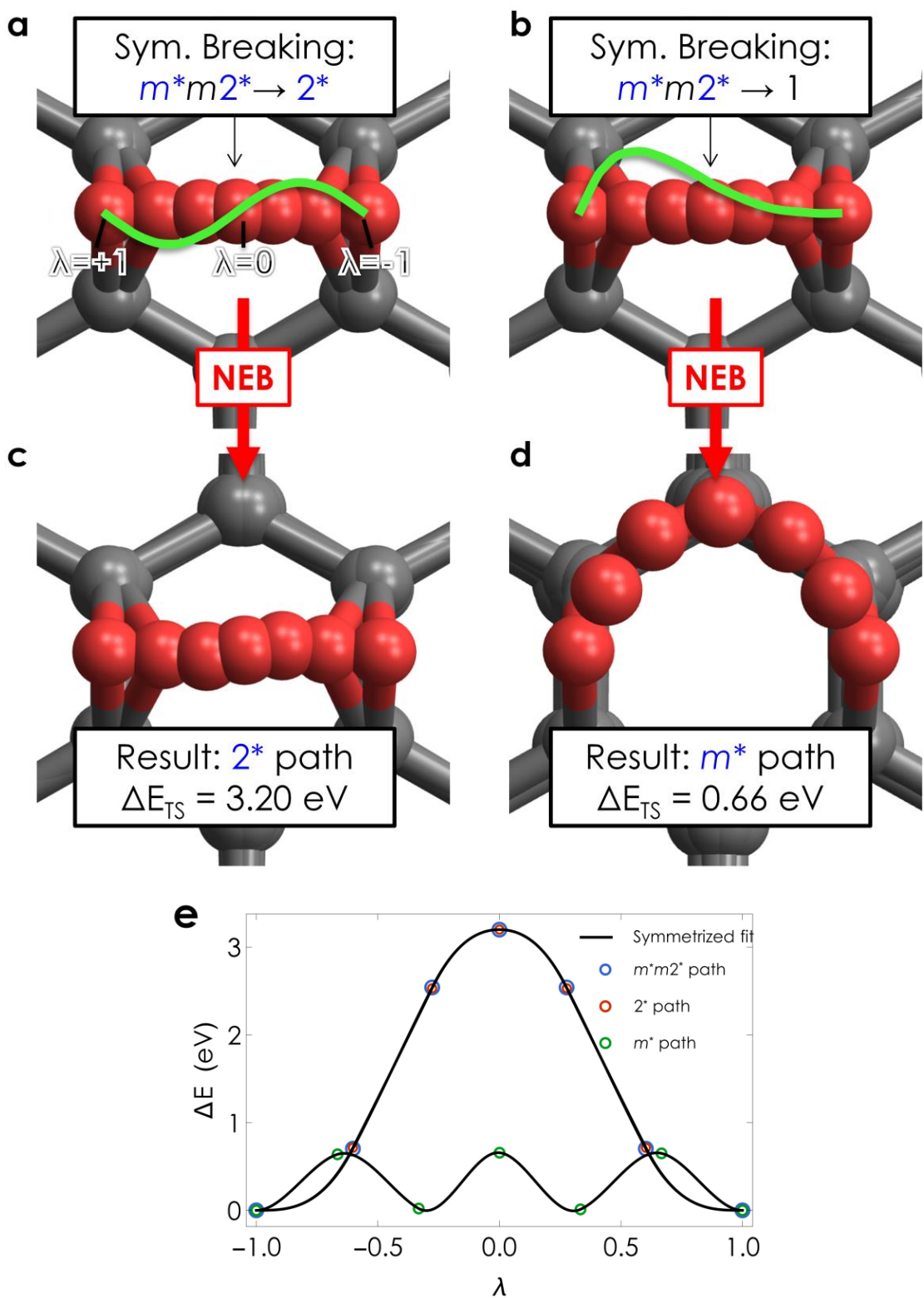

**Figure 3: The consequences of distortion symmetry and balanced forces for NEB calculations.** Superimposed images along oxygen diffusion paths on graphene after NEB. Starting from a $m^*m2^*$ path that goes directly across a $C_6$ ring, perturbations were added with $2^*$ symmetry in **a** and trivial symmetry in **b**. These perturbations are exaggerated; the maximum displacement of oxygen along the path was 0.1 Å in **a** and about 0.18 Å in **b**. **c** and **d** show the final paths after NEB relaxation starting from the perturbed paths of **a** and **b** respectively. **e** shows the calculated energies of the images and the interpolation provided by Quantum Espresso's NEB module[35]. More details are given in the Supplementary Discussion 2.

Figure 3e shows that the path with trivial symmetry (*i.e.* point group 1) relaxes to a much lower energy path with *m\** symmetry. Because NEB can only raise the symmetry of the path, not lower it (see Methods), the *2\** path cannot achieve the same results and has approximately the same energy as the original relaxed *m\*m2\** path. Such an analysis is applicable to many other types of problems such as the MEPs for ferroelectric and magnetic switching of domains[36]. Essentially the same path as our *m\*m2\** path was studied by Dai et al.[37] who reported a high-energy transition state with a barrier of 1.75 eV. Dai et al. also report a lower energy transition state, apparently similar to our 0.66 eV state, but with 0.81 eV and an energy profile that is highly asymmetric with respect to the distortion coordinate and thus erroneously violates the *m\** symmetry that our symmetry analysis in Figure 3 indicates it must possess. Such unintentional errors are in fact quite common in literature as the survey examples in Supplementary Table 1 indicate. Supplementary Table 1 gives about 50 examples of published studies where distortion symmetry would have been useful; this is a very small subset of such studies. We note that, although in conventional chemical reactions where products and reactants are not energetically equivalent there are no distortion-reversing symmetry operations (i.e. starred operations), our survey suggests that most references to the use of NEB in materials science are made in the study of pathways with energetically equivalent endpoints, such as in diffusion, dislocation, domain wall, interface, and grain boundary motion.

Next we demonstrate the application of distortion groups in predicting allowed energy couplings that are odd powers in the distortion parameter and may appear at first to be disallowed by conventional symmetry groups. We will use beta barium borate, $\beta$-BaB$_2$O$_4$, a widely used nonlinear optical crystal, as an example. Using a parent structure ($\lambda$=0) with $R\bar{3}c$ symmetry[38], we construct a distortion with $R3c$ variants at $\lambda$=-1 and $\lambda$=+1, where the displacements scale linearly with $\lambda$. This distortion pathway has $R\bar{3}^*c$ symmetry. The calculated energy profile, $\Delta E(\lambda)$, is shown in Fig. 4a and is symmetric with respect to $\lambda$. This is a consequence of the starred symmetry operations, just as with the PF$_5$ example. In Figure 4b, we depict the sequence of intermediate structures along the distortion pathway by superimposing them using a color scale. From the blurred pattern, we can see that this distortion is mostly the nearly rigid rotation of the B$_3$O$_6$ rings. For $\beta$-BaB$_2$O$_4$ and distortion group

$R\bar{3}^*c$ (No. 4306 in the complete double antisymmetry space group (DASG) listing[39,40]), the $B_3O_6$ rings are on the 12c site. From referring to the listing, this means that there are rings located at {0,0,z}, {0,0,-z+½}, {0,0,-z}, and {0,0,z+½} with rotation vectors of [0,0,$\omega_z$], [0,0, $\omega_z$], [0,0,-$\omega_z$], and [0,0,-$\omega_z$] respectively. This tells us that the $R\bar{3}^*c$ symmetry requires that the rotation ($\omega$) of the rings is only along the z-axis and alternates every two rings along the column, *i.e.* clockwise, clockwise, counterclockwise, counterclockwise, *etc*. The distortion symmetry listing also tells us that the displacement of the rings is only along the z axis and all the rings displace in the same direction with the same magnitude. This is just one of the many ways in which the concept of distortion symmetry can be used to make useful predictions about distortions.

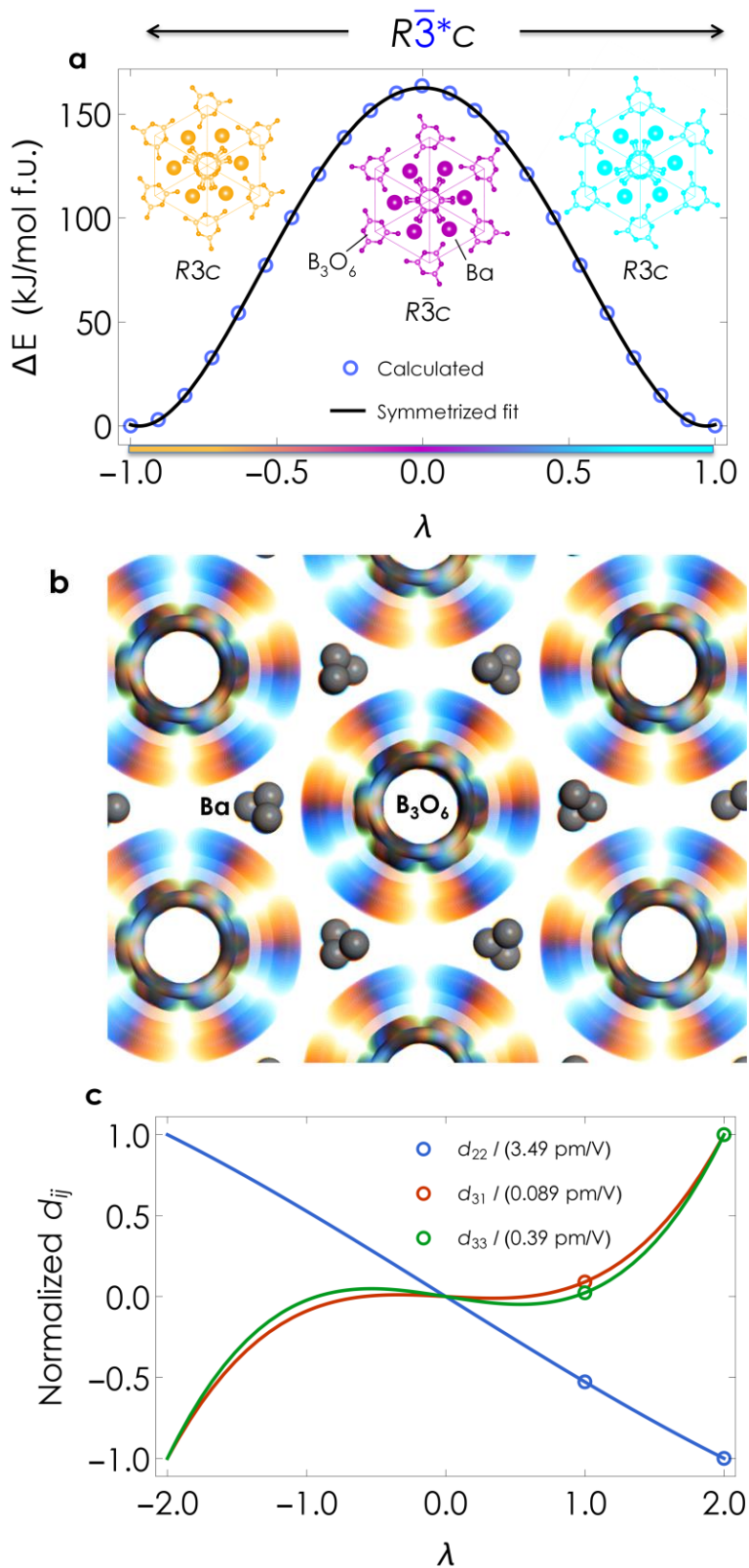

**Figure 4: The application of distortion symmetry to a distortion of β-BaB$_2$O$_4$.** The energy profile in **a** is symmetric due to the starred symmetry. The structures at λ=-1, 0, and +1 are depicted in orange, magenta, and cyan respectively. **b** shows the superimposed images of β-BaB$_2$O$_4$ along the distortion pathway from λ=-1 to +1; their color varies from orange to cyan as λ varies from -1 to +1. **c** shows the SHG coefficients along this pathway calculated by Cammarata & Rondinelli[38] and a polynomial fit using only the coefficients that are consistent with $\bar{3}^*m$ symmetry.

It is certainly not intuitive how properties, such as optical second harmonic generation, relevant to this material would vary with this distortion. The SHG interaction, $P_i^{2\omega} = d_{ijk} E_j^\omega E_k^\omega$, creates a nonlinear polarization $P$ at a frequency of $2\omega$ by combining two photons with electric fields $E$ at frequency $\omega$. In Fig. 4c, we plot the calculated values for optical second harmonic generation (SHG) coefficients for this crystal along the distortion pathway as calculated by Cammarata & Rondinelli[38]. The macroscopic point group of the β-BaB$_2$O$_4$ distortion described above is $\bar{3}^*m$. We write $d_{ijk}$ as a function of $\lambda$ as,

$$d_{ijk}(\lambda) = A_{ijk} + B_{ijk}\lambda + C_{ijk}\lambda^2 + D_{ijk}\lambda^3 + \cdots \qquad (2)$$

$\bar{1}^*$, an element of $\bar{3}^*m$, combined with Neumann's principle, requires that $d_{ijk}(\lambda) = -d_{ijk}(-\lambda)$. Thus we immediately deduce that the function should be odd with respect to $\lambda$, and hence the even coefficients (A, C, etc.) should be exactly zero. It also clearly implies that $d_{ijk}(0)=0$. The points marked by open circles at $\lambda=1.0$ and $\lambda=2.0$ are from previously reported calculations[38]. The curves are the result of solving for $d_{ijk}(\lambda)=B_{ijk}\lambda+D_{ijk}\lambda^3$ that goes through these points. Since it was not obvious *a priori* that $d_{ijk}$ should be an odd function of this distortion, this example demonstrates how applying distortion symmetry predicts the form of the tensors that describe the change in *any* property as a function of a distortion. This also suggests that in the Landau phenomenology, there should be an energy coupling of the form

$$U \propto Q_{ijk} P_i^{2\omega} P_j^\omega P_k^\omega \lambda + R_{ijk} P_i^{2\omega} P_j^\omega P_k^\omega \lambda^3 \qquad (3)$$

in the parent $R\bar{3}c$ (point group $\bar{3}m$) phase. However, the polar third rank tensors $Q_{ijk}$ and $R_{ijk}$ would be identically zero in the conventional $\bar{3}m$ parent phase as deduced by noticing that $\bar{1}(P)=-P$ and $\bar{1}(\lambda)=\lambda$. The only way such a coupling would exist is if the complete distortion symmetry of the pathway, namely, $\bar{3}^*m$ is considered, since $\bar{1}^*(P)=-P$ and $\bar{1}^*(\lambda)=-\lambda$. This example shows the value of distortion symmetry analysis in easily revealing energy invariants that are odd powers in λ. It is much more transparent than the corresponding representation theory based analysis.

Including distortion-reversing symmetry operations (*i.e.* starred operations) can place greater restrictions on invariant property tensors. As seen from Table 1 which compares the form of various types of tensors for a "conventional" symmetry group versus a group includes starred operations. Because of how the $A_{ijk}$ and $C_{ijk}\lambda^2$ terms transform, $A_{ijk}$ and $C_{ijk}$ are 1*-even 3rd rank polar tensors. Likewise, $B_{ijk}$ and $D_{ijk}$ are 1*-odd 3rd rank polar tensors. From consulting Table 1, we find that the power series expansion of $d_{ijk}(\lambda)$ to the third power contains half as many degrees of freedom if the full symmetry group $\bar{3}^*m$ is considered instead of only the unstarred symmetry of the distortion, $3m$. If instead of the $P_i^{2\omega} = d_{ijk}E_j^\omega E_k^\omega$ interaction, we had considered $P_i^{2\omega} = d_{ijk}E_j^\omega (\nabla \times E)_k^\omega$ as an example, then the corresponding tensors $A_{ijk}$ and $C_{ijk}$ in an expansion similar to (2) would be axial 1*-even, 1'-even tensors, while the tensors $B_{ijk}$ and $D_{ijk}$ would be axial, 1*-odd, 1'-even tensors. Their forms and how they differ between the distortion group and conventional group is also given in Table 1. Thus, distortion symmetry can significantly reduce the number of tensor coefficients that are predicted to be non-zero.

## Table 1

Comparison between general $\bar{3}^*m$-invariant tensors (green) and $3m$-invariant tensors (red) for selected tensor types.

| | 0$^{th}$ rank<br>$Q$<br>(1) | 1$^{st}$ rank<br>$Q_i$<br>(V) | 2$^{nd}$ rank<br>$Q_{ij}$<br>(V$^2$) | 2$^{nd}$ rank<br>$Q_{ij}=Q_{ji}$<br>([V]$^2$) | 3$^{rd}$ rank<br>$Q_{ijk}=Q_{ikj}$<br>(V[V]$^2$) | 4$^{th}$ rank<br>$Q_{ijkl}=Q_{jikl}=Q_{klij}$<br>([[V]$^2$]$^2$) |
|---|---|---|---|---|---|---|
| Polar,<br>1'-even,<br>1*-even<br>(1) | $Q$ | $\begin{pmatrix} 0 \\ 0 \\ 0\neq Q_3 \end{pmatrix}$ | $\begin{pmatrix} Q_{11} & 0 & 0 \\ 0 & Q_{11} & 0 \\ 0 & 0 & Q_{33} \end{pmatrix}$ | $\begin{pmatrix} Q_{11} & 0 & 0 \\ 0 & Q_{11} & 0 \\ 0 & 0 & Q_{33} \end{pmatrix}$ | $\begin{pmatrix} 0 & 0 & 0 & 0 & 0\neq Q_{15} & 0\neq Q_{16} \\ 0\neq Q_{16} & 0\neq -Q_{16} & 0 & 0\neq Q_{15} & 0 & 0 \\ 0\neq Q_{31} & 0\neq Q_{31} & 0\neq Q_{33} & 0 & 0 & 0 \end{pmatrix}$ | $\begin{pmatrix} Q_{11} & Q_{12} & Q_{13} & Q_{14} & 0 & 0 \\ Q_{12} & Q_{11} & Q_{13} & -Q_{14} & 0 & 0 \\ Q_{13} & Q_{13} & Q_{33} & 0 & 0 & 0 \\ Q_{14} & -Q_{14} & 0 & Q_{55} & 0 & 0 \\ 0 & 0 & 0 & 0 & Q_{55} & Q_{14} \\ 0 & 0 & 0 & 0 & Q_{14} & \frac{Q_{11}-Q_{12}}{2} \end{pmatrix}$ |
| Axial,<br>1'-even,<br>1*-even<br>(e) | 0 | $\begin{pmatrix} 0 \\ 0 \\ 0 \end{pmatrix}$ | $\begin{pmatrix} 0 & 0\neq Q_{12} & 0 \\ 0\neq -Q_{12} & 0 & 0 \\ 0 & 0 & 0 \end{pmatrix}$ | $\begin{pmatrix} 0 & 0 & 0 \\ 0 & 0 & 0 \\ 0 & 0 & 0 \end{pmatrix}$ | $\begin{pmatrix} Q_{11} & -Q_{11} & 0 & Q_{14} & 0 & 0 \\ 0 & 0 & 0 & 0 & -Q_{14} & -Q_{11} \\ 0 & 0 & 0 & 0 & 0 & 0 \end{pmatrix}$ | $\begin{pmatrix} 0 & 0 & 0 & 0 & 0\neq Q_{15} & 0 \\ 0 & 0 & 0 & 0 & 0\neq -Q_{15} & 0 \\ 0 & 0 & 0 & 0 & 0 & 0 \\ 0 & 0 & 0 & 0 & 0 & 0\neq -Q_{15} \\ 0\neq Q_{15} & 0\neq -Q_{15} & 0 & 0 & 0 & 0 \\ 0 & 0 & 0 & 0\neq -Q_{15} & 0 & 0 \end{pmatrix}$ |
| Polar,<br>1'-even,<br>1*-odd<br>(b) | $0\neq Q$ | $\begin{pmatrix} 0 \\ 0 \\ Q_3 \end{pmatrix}$ | $\begin{pmatrix} 0\neq Q_{11} & 0 & 0 \\ 0 & 0\neq Q_{11} & 0 \\ 0 & 0 & 0\neq Q_{33} \end{pmatrix}$ | $\begin{pmatrix} 0\neq Q_{11} & 0 & 0 \\ 0 & 0\neq Q_{11} & 0 \\ 0 & 0 & 0\neq Q_{33} \end{pmatrix}$ | $\begin{pmatrix} 0 & 0 & 0 & 0 & Q_{15} & Q_{16} \\ Q_{16} & -Q_{16} & 0 & Q_{15} & 0 & 0 \\ Q_{31} & Q_{31} & Q_{33} & 0 & 0 & 0 \end{pmatrix}$ | $\begin{pmatrix} 0\neq Q_{11} & 0\neq Q_{12} & 0\neq Q_{13} & 0\neq Q_{14} & 0 & 0 \\ 0\neq Q_{12} & 0\neq Q_{11} & 0\neq Q_{13} & 0\neq -Q_{14} & 0 & 0 \\ 0\neq Q_{13} & 0\neq Q_{13} & 0\neq Q_{33} & 0 & 0 & 0 \\ 0\neq Q_{14} & 0\neq -Q_{14} & 0 & 0\neq Q_{55} & 0 & 0 \\ 0 & 0 & 0 & 0 & 0\neq Q_{55} & 0\neq Q_{14} \\ 0 & 0 & 0 & 0 & 0\neq Q_{14} & 0\neq \frac{Q_{11}-Q_{12}}{2} \end{pmatrix}$ |
| Axial,<br>1'-even,<br>1*-odd<br>(be) | 0 | $\begin{pmatrix} 0 \\ 0 \\ 0 \end{pmatrix}$ | $\begin{pmatrix} 0 & Q_{12} & 0 \\ -Q_{12} & 0 & 0 \\ 0 & 0 & 0 \end{pmatrix}$ | $\begin{pmatrix} 0 & 0 & 0 \\ 0 & 0 & 0 \\ 0 & 0 & 0 \end{pmatrix}$ | $\begin{pmatrix} 0\neq Q_{11} & 0\neq -Q_{11} & 0 & 0\neq Q_{14} & 0 & 0 \\ 0 & 0 & 0 & 0 & 0\neq -Q_{14} & 0\neq -Q_{11} \\ 0 & 0 & 0 & 0 & 0 & 0 \end{pmatrix}$ | $\begin{pmatrix} 0 & 0 & 0 & 0 & Q_{15} & 0 \\ 0 & 0 & 0 & 0 & -Q_{15} & 0 \\ 0 & 0 & 0 & 0 & 0 & 0 \\ 0 & 0 & 0 & 0 & 0 & -Q_{15} \\ Q_{15} & -Q_{15} & 0 & 0 & 0 & 0 \\ 0 & 0 & 0 & -Q_{15} & 0 & 0 \end{pmatrix}$ |

The ubiquitousness of distortion symmetry is further illustrated in Figure 5 with four examples. Each panel depicts the structures from $\lambda=-1$ to $\lambda=+1$ superimposed so that the movement of the atoms appears in the form of a blur. A common piezoelectric crystal quartz ($SiO_2$) is depicted in Fig. 5a, where a distortion from one domain of alpha quartz (at $\lambda=-1$) through beta quartz (at $\lambda=0$) to another domain of alpha (at $\lambda=+1$) exhibits the distortion symmetry of $P6_4*22*$ (with point group $6*22*$). Supplementary Discussion 3 shows how there is an equivalent pathway in left-handed quartz with $P6_2*22*$ symmetry, as well as the symmetries of paths between left and right handed quartz. A prototypical proper ferroelectric, $PbTiO_3$ is depicted in Fig. 5b, where the distortion pathway runs between opposite polarization states and has $P4/m*mm$ symmetry. An improper ferroelectric antiferromagnet, $YMnO_3$, distorting from one ferroelectric domain, $\alpha^+$ at $\lambda=-1$ to the opposite domain $\alpha^-$ at $\lambda=+1$ exhibits a distortion symmetry of $P6_3/m*cm$. Note that the corresponding point groups ($4/m*mm$ and $6/m*mm$) for the panels a and b allow for an energy invariant of the form $U \propto P.\lambda + P.\lambda^3 + ...$, where $P$ is the polarization that develops under the distortion modes in question, parametrized by $\lambda$. In contrast, this coupling is zero under the conventional parent phase symmetries of $m\bar{3}m$ and $6/mmm$, respectively, again demonstrating the value of the distortion reversal symmetry in revealing such couplings in a transparent and simple manner. This coupling in $YMnO_3$ was confirmed by first principles calculations[41]. Including antiferromagnetism and weak canted ferromagnetism in $YMnO_3$[42], we can consider two cases: either spins reverse or spins are invariant through $\alpha^+ \to \alpha^-$. The former has $P6_3'/m*$ symmetry and the latter has $P6_3'/m'*$ symmetry. Note that these double antisymmetry space groups involve two independent antisymmetries, $1*$ and $1'$. A complete listing of the 17,803 double antisymmetry space groups has recently been made available by VanLeeuwen et al.[39,40]. These kinds of distortion pathways should exist in most ferroelectrics and multiferroics. The 670 cm$^{-1}$ $B_{1u}$ mode of a superconductor, $YBa_2Cu_3O_{6.5}$, is shown in Fig. 5d[19]. This mode has a distortion symmetry of

*Pm\*mm*, and has recently been shown to couple with $A_g$ modes to create a transient structure that was reported to exhibit room temperature coherent interlayer transport on picosecond time scales, reminiscent of superconductivity[20,21]. Including the coupling between this $B_{1u}$ mode and the $A_g$ modes retains the same distortion symmetry. The form of the invariant tensors describing changes in *any* property in these example systems as a function of distortion can be deduced, similar to that shown in Table 1.

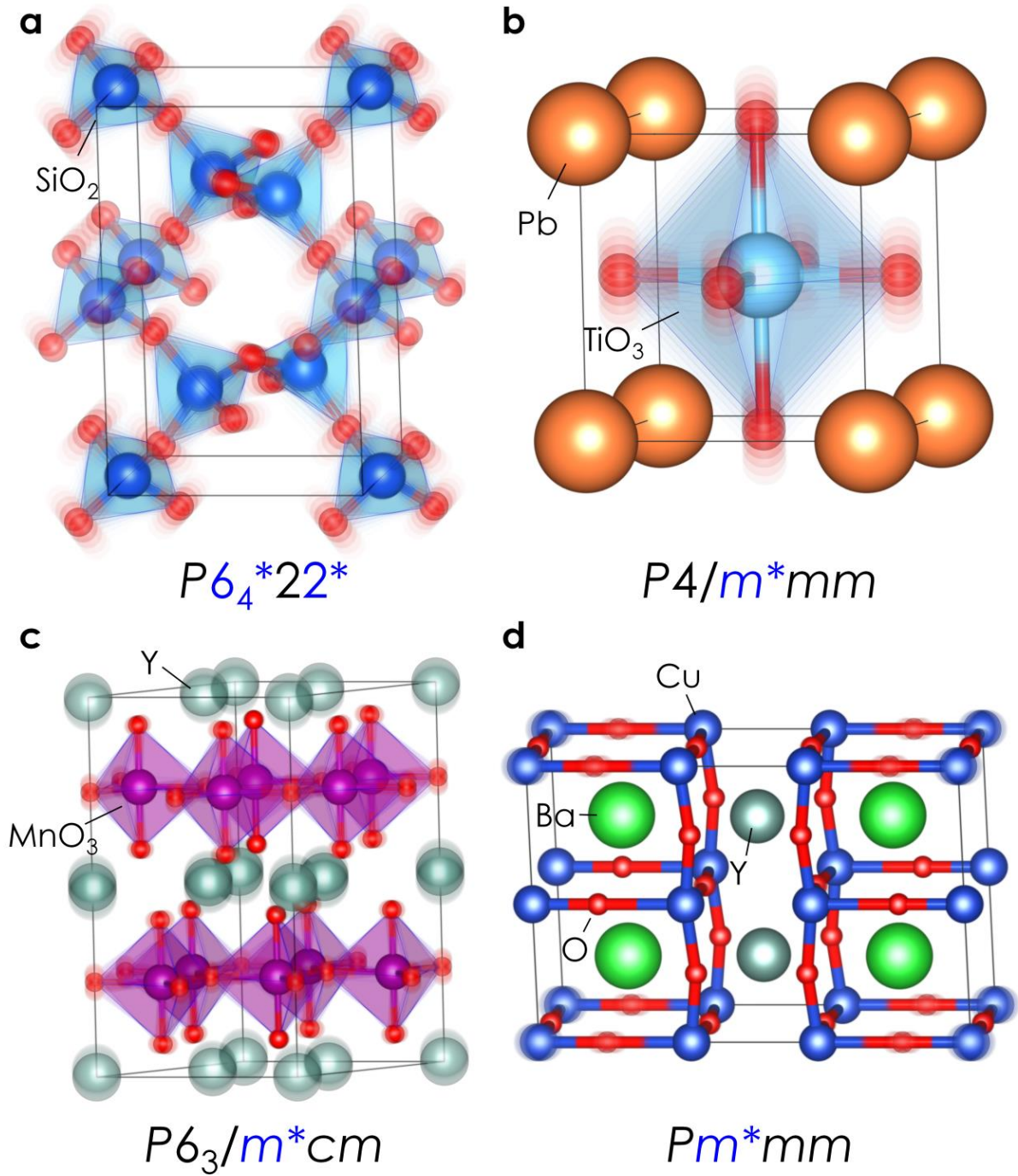

**Figure 5: Four different distortions in crystals and their distortion symmetry groups.** **a** shows a distortion pathway between two domain variants of alpha quartz, passing through beta quartz at $\lambda=0$. **b** shows a distortion of $PbTiO_3$ created by linearly interpolation between opposite polarization states. **c** shows a distortion of $YMnO_3$ between opposite domain variants. **d** shows a $B_{1u}$ normal mode of $YBa_2Cu_3O_{6.5}$.

Finally, we note that distortion symmetry can be applied to the electronic structure of a distortion and has implications for Berry phase calculations. In the Supplementary Discussion 5, we show that $1^*\gamma = -\gamma$ where $\gamma$ is the Berry phase calculated for the path of the distortion and assuming $|\gamma| < \pi$. Thus, if a distortion path is invariant with $R^*$ symmetry (i.e. if $R^*\gamma = \gamma$), then we conclude that

$$\gamma = 0. \qquad (4)$$

This general result states that for *any* distortion pathway with a starred symmetry, the Berry phase will be *exactly* zero. Ceresoli & Tosatti[43] using Berry phase to compute the orbital magnetic moment for a pseudorotation distortion of benzene. They report their computation of the magnetic screening factor, $\sigma$, which is proportional to the Berry phase $\gamma$, as follows: "The nearly exact balance of positive and negative currents explains the globally small magnetic screening in benzene." We have determined that this distortion with in-plane displacements of atoms in the benzene plane has $m^*m^*m$ type symmetry and, by the arguments presented above, therefore should have "exactly" zero magnetic screening.

In the course of this study, we have come to conclude that distortion symmetry introduced here has applicability to a very wide range of physical problems. Besides the examples presented above, the symmetry could be applicable to reconfigurations of proteins and other biomolecules, motion of domain walls and grain boundaries, distortion tuning of metamaterials such as those exhibiting photonic bandgaps[44], distortion-reversal symmetry protection of topological boundary modes[45] analogous to time-reversal symmetry protection of topological insulators via Kramers theorem, the search for transient and metastable phases exhibiting emergent properties under a distortion[19–24], and the search for intermediate stable structures in materials away from equilibrium, by reducing the asymmetric domain. The double antisymmetry groups created from incorporating both distortion-reversal and time-reversal symmetries could be applied to explore the energy landscape of magnetic structures under a distortion. Similar to the impact of time-reversal antisymmetry and magnetic groups, we foresee a commensurate impact of distortion-reversal antisymmetry and distortion groups on a diverse set of problems and design tools used in the physical sciences.

## Methods:

**The decomposition seen in Fig. 1 and the basis of Supplementary Fig. 1**

The decomposition applied to the simple distortion seen in Fig. 1 was performed by selecting a basis for translation, rotation, scaling, and decomposition components. Supplementary Fig. 1 shows an example of such a basis and, in this case, it is also symmetry-adapted. After the basis is selected, the displacement vectors are projected onto it and each component can be isolated as shown in Fig. 1. This and other notions of decomposition into rotational and non-rotational components were explored in our attempts to formalize the concepts of rotation-reversal symmetry.

**$PF_5$ Pseudorotation Nudged Elastic Band (NEB) Calculations**

The $PF_5$ pseudorotation minimum energy pathway was computed using $DMol^3$ in Materials Studio 6.0[32]. The approximate structure was input using Materials Studio's tools and then geometrically optimized using $DMol^3$. This structure was taken as the $\lambda=-1$ variant and was copied and rotated 90° around the high symmetry axis to make the $\lambda=+1$ variant. Then the Reaction Preview tool was used to match the atoms of the structures and generate an initial guess path. This guess was used as input for the $DMol^3$ Transition State Search tool whose output was then run in the Transition State Conformation Tool which performs using NEB to find a minimum energy pathway. The output from NEB was symmetrized to remove the small asymmetric numerical errors and used to construct the plots in Fig. 2.

**Oxygen diffusion on graphene NEB Calculations**

The geometrically optimized structure was from an example calculation used at the QE2014 workshop held at PennState. The $\lambda=-1$ structure consists of a 3x3 supercell of graphene with an oxygen atom bonded to the surface, as part of an epoxy functional group. This was mirrored to create the $\lambda=+1$ structure. These structures were used as the first and final images in the input for Quantum Espresso's NEB module (neb.x)[35]. Seven images were used. These are linearly interpolated from the first and final images. This initial guess path, discretized into a chain of seven images, relaxed into the path seen in Fig. 3a and Fig. 3b with *m*m2* symmetry.

Next, two new paths were created from the *m*m2* path using small symmetry breaking perturbations of the oxygen trajectory parallel to the graphene sheet. The first was a sinusoidal perturbation with an amplitude of 0.1 Å resulting in a path with 2* symmetry. The second was a perturbation of $-(\lambda^5 - 5\lambda^4 - 6\lambda^3 + 2\lambda^2 + 7\lambda + 3)/32$Å resulting in a path with only trivial symmetry (*i.e.* point group 1). These two new paths were then relaxed using QE's NEB module again to get the paths shown in Fig. 3c and Fig. 3d.

The reason that starred symmetry operations affect the results of NEB calculations in this way is because NEB commutes with 1* in the same way that conventional symmetry operations commute with physical laws. Clearly, NEB(X) gives the same result as $A^{-1}$NEB(AX) where X is the initial guess path and A is an ordinary symmetry operation, such as a rotation or a mirror. Similarly, NEB(X) gives the same result as $1^{*-1}$NEB(1*X) and, since $1^{*-1}$=1*, 1*NEB(1*X). This is no different from the idea that physical laws should not depend on what basis one chooses for their coordinate system. If A* is a symmetry of the initial guess path, *i.e.* X = A*X, then, by substitution, NEB(X) = A*NEB(X). Thus the commutativity of A* with NEB guarantees that X = A*X implies NEB(X) = A*NEB(X), i.e. that a symmetry of the initial guess will also be a symmetry of the results. In practice, however, $A^{-1}NEB(AX)$ is not exactly equal to NEB(X) because the NEB implementation will have small symmetry breaking numerical errors.

**$\beta$-BaB$_2$O$_4$ (BBO) Calculations**

The $\beta$-BaB$_2$O$_4$ distortion shown in Fig. 4 was created using Materials Studio's Reaction Preview tool by matching atoms of a $\beta$-BaB$_2$O$_4$ variant with its inverted variant. The result is a path that goes through an $R\bar{3}c$ intermediate, as shown in Fig. 4a and Fig. 4b. The energy along this path, as plotted in Fig. 4a, was computed using Materials Studio's CASTEP module and symmetrized to remove small asymmetric numerical errors. Similar methods were applied to make the energy plot for the quartz example in Supplementary Fig. 1.

Our $\beta$-BaB$_2$O$_4$ distortion path is similar, but not identical, to the linear path implied by Cammarata & Rondinelli[38] where the displacements from the hypothetical $R\bar{3}c$ parent structure are scaled by a factor. In particular, we note that our path has rigid or near rigid rotation of the B$_3$O$_6$ rings whereas linearly scaling the displacements creates a path that diverges from rigid rotation as rotation angle increases. Nonetheless, as the two paths are still very similar and have the same distortion symmetry, so we have used the results of Cammarata & Rondinelli[38] to create Fig. 4c.

**Determining distortion symmetry group**

Let $S(\lambda)$ denote the structure at $\lambda$. Let $G(\lambda)$ denote the conventional symmetry group of $S(\lambda)$. If there is exists $A \in G(\lambda = 0)$ such that $AS(\lambda) = S(-\lambda)$ for all $-1 \leq \lambda \leq +1$, then the symmetry of the distortion is $H \cup 1^*AH$ where $H = \cap_{-1 \leq \lambda \leq +1} G(\lambda)$. Otherwise, $H$ is the symmetry of the distortion. In other words, find the conventional symmetry group of all the images in a pathway ($\lambda$ from -1 to +1); the intersection of these groups is the group $H$. Now find an element "A" in the conventional symmetry group of the structure at $\lambda = 0$ that can transform a structure at $\lambda$ to a structure at $-\lambda$. The distortion group of the pathway is then $H \cup A^*H$. If no such $A$ can be found, then $H$ is the symmetry of the distortion pathway.

Acknowledgements:

The authors acknowledge financial support from the Penn State MRSEC Center for Nanoscale Science through the National Science Foundation grant numbers DMR-1420620 & 0820404, and National Science Foundation grant DMR-1210588. Discussions with M. Huang, D. B. Litvin, V. H. Crespi, M. I. Aroyo, B. J. Campbell, I. Dabo, and C. X. Liu are gratefully acknowledged. J. M. Rondinelli provided valuable information for Figure 4 and related discussion.

Author Contributions:

BKV found the initial problems with VG's rotation-reversal symmetry and distortion symmetry evolved out of the discussions with VG about these problems. BKV contributed the idea of replacing polyhedral rotations with the general parameterization of a pathway by $\lambda$, 1* for reversing $\lambda$, and the application to NEB and transition state theory. VG contributed sections on applying Neumann's principle, Landau phenomenology (energy couplings and energy invariants), initial ideas on the decomposition seen in Fig. 1, and proposed studying the distortions seen in Figures 4, 5, and Supplementary Figure 1. BKV and VG co-wrote the paper and made the figures. Computational results and Supplementary Table 1 are from BKV.

Competing Financial Interests statement:

The authors declare no competing financial interests.

## Supplementary Discussion 1: Connection between distortion symmetry and representation analysis

We will further establish the connection between distortion symmetry and representation analysis with the example of atomic displacements of a water molecule, $H_2O$, with conventional symmetry $mm2$ ($C_{2v}$) as depicted in Supplementary Fig. 1a. Using the irreducible representations of $mm2$, we can construct a symmetry-adapted basis for the atomic displacements of $H_2O$. Our chosen basis is depicted in Supplementary Fig. 1a. Each of our basis modes correspond to a distortion path that is constructed by linearly scaling the displacements by $\lambda$, as depicted in Fig. 1h. The irrep carried by the span of each mode in our basis is labeled along with the symmetry of the corresponding distortion. The $A_1$ distortions have $mm2$ symmetry. The $A_2$, $B_1$, and $B_2$ distortions have $m*m*2$, $m*m2*$, and $mm*2*$ symmetries respectively. The properties of $mm2$ allow for this simple correspondence between irreps and distortion groups. In general though, such correspondences can be more complex when 2- or higher-dimensional irreps exist. For example, the ammonia molecule, $NH_3$, has $3m$ symmetry. $3m$ has a 2-dimensional irrep, denoted E. This irrep is carried by the linear span of $v_1$ and $v_2$ shown in Supplementary Figure 1b. If we construct distortions by linearly scaling $v_1$ and $v_2$ by $\lambda$, then the resulting distortions have symmetry $m$ and $m*$ respectively. Thus, knowing the irreducible components of a set of displacements is not equivalent to knowing the symmetry of the corresponding distortion. The symmetry of distortions constructed from linear combinations of $v_1$ and $v_2$ is shown in Supplementary Fig. 1c. While most of this space (orange region) does have the kernel symmetry of 1, one can see specific linear combinations of $v_1$ and $v_2$ (black lines) which give rise to $m$ and $m*$ symmetries. Complexities of representation analysis, such as in 2- or higher-dimensional irreps, are avoided by using distortion symmetry.

For many problems, distortion symmetry offers a simple and elegant alternative to traditional representation analysis.

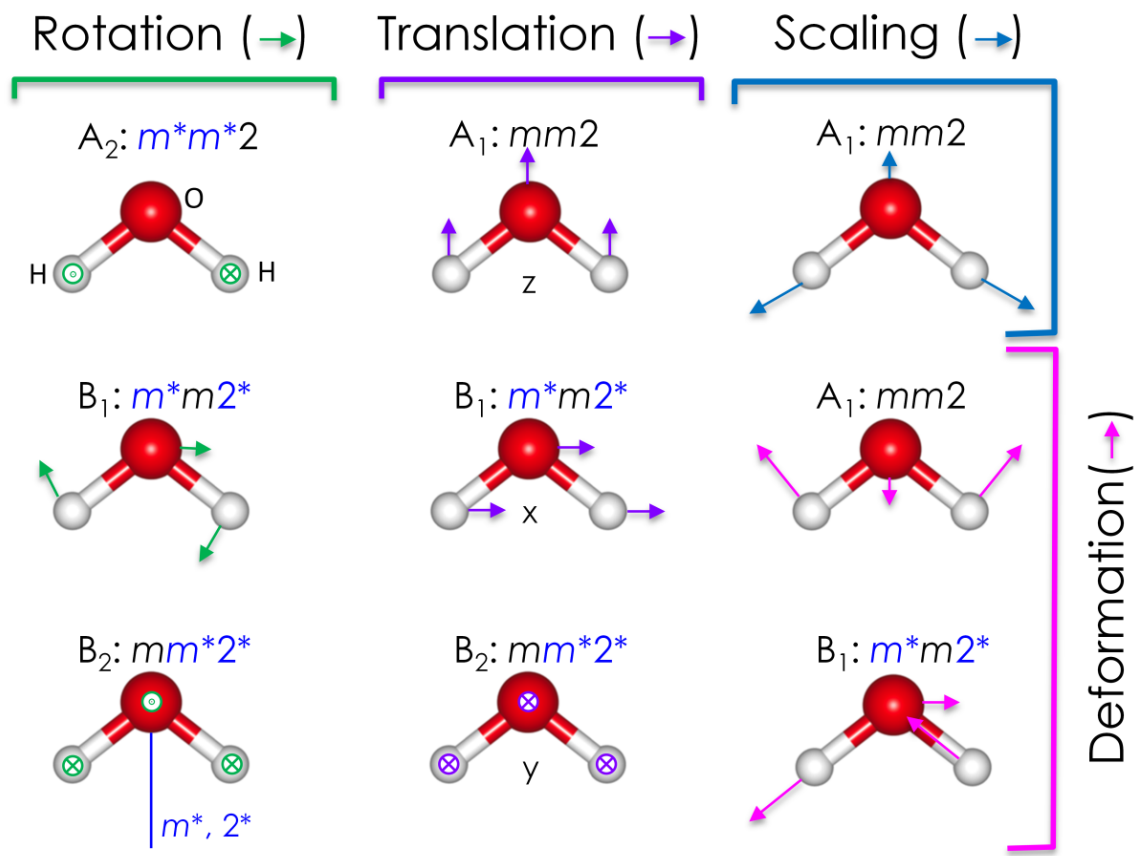
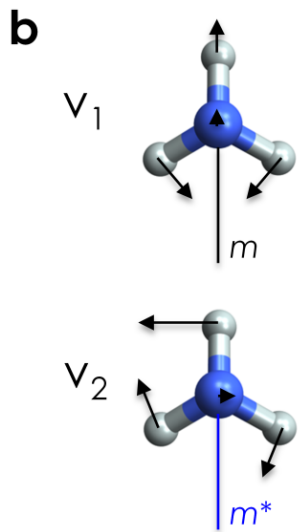
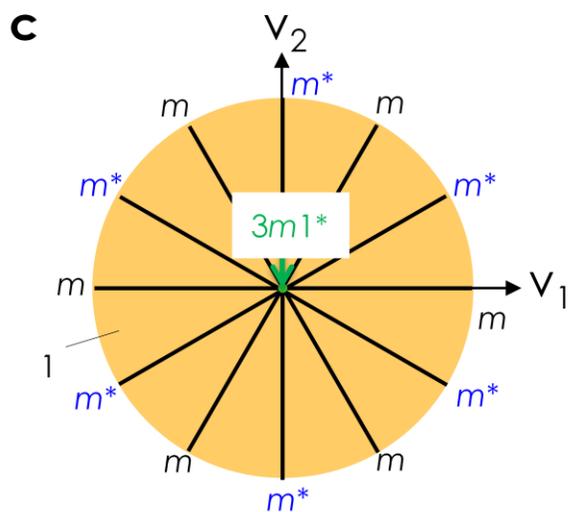

**Supplementary Figure 1: A comparison of representation analysis and distortion symmetry for simple molecules.** A symmetry adapted basis is given in **a** for the atomic displacements of a water molecule. Displacement vectors $v_1$ and $v_2$ for NH$_3$ and their distortion symmetry in **b**. The symmetry of linear combinations of $v_1$ and $v_2$ in **c**.

An improper ferroelectric antiferromagnet, YMnO$_3$, distorting from one ferroelectric domain, $a^+$ at $\lambda=-1$ to the opposite domain $a^-$ at $\lambda=+1$ exhibits a distortion symmetry of $P6_3/m^*cm$ (Fig. 5c). This is also effectively the distortion implied by Fennie and Rabe[1] in studying the $P6_3/mmc$ parent structure. It is also an interesting case in terms of the relationship between distortion symmetry and representation analysis. Fennie and Rabe identify an unstable K$_3$ phonon mode of the $P6_3/mmc$ parent structure as driving the improper ferroelectric transition. K$_3$ is a 2D irrep and thus corresponds to two order parameters. Depending on order parameter direction, a perturbation that transforms as K$_3$ could result in $P6_3cm$, $P\bar{3}c1$, or $P3c1$ type symmetry. $P6_3cm$ and $P\bar{3}c1$ correspond to 1D subspaces of K$_3$ and their transformation properties should require that the corresponding distortions have $P6_3/m^*cm$ and $P6_3^*/m^*cm^*$ symmetry respectively. $P3c1$ is the kernel, the lowest possible symmetry achievable with a perturbation that transforms as K$_3$; it is the symmetry of a general point in the 2D order parameter space of K$_3$. Distortions corresponding to these points should have $P\bar{6}^*c2^*$ symmetry. In the YMnO$_3$ case, the order parameter direction corresponds to $P6_3/m^*cm$ and the coupled modes should not change this. We note the considerable complexity of describing the symmetry of this distortion with representation analysis (*i.e.* the last few sentences) versus comparative simplicity of the distortion symmetry classification (*i.e.* $P6_3/m^*cm$). This is an advantage of distortion symmetry over conventional representation analysis. We also note that giving the primary irrep of a distortion is *not* equivalent to giving a distortion group, just as is the case with magnetic symmetry[2], because K$_3$ actually corresponds to three different types of distortion symmetry: $P6_3/m^*cm$, $P6_3^*/m^*cm^*$, and $P\bar{6}^*c2^*$.

Finally, we note one way in which the representations of distortion groups can be applied. The path depicted in Supplementary Fig. 2 has *m*m2* symmetry. The following is a *m*m2* character table:

|    | 1  | 2* | m  | m* | Kernel |
|----|----|----|----|----|--------|
| Γ₁ | 1  | 1  | 1  | 1  | *m*m2* |
| Γ₂ | 1  | 1  | -1 | -1 | 2*     |
| Γ₃ | 1  | -1 | 1  | -1 | m      |
| Γ₄ | 1  | -1 | -1 | 1  | m*     |

If we consider perturbations that displace the oxygen atom and not the carbon atoms, the 21 dimensional space of perturbations of the path carry a representation of *m*m2* with the following irreducible components: 7 $\Gamma_1$ + 3 $\Gamma_2$ + 7 $\Gamma_3$ + 4 $\Gamma_4$. Using a symmetry-adapted basis, this can be decomposed into four symmetry invariant subspaces:

- the span of {Δx(-1)-Δx(1), Δz(-1)+Δz(1), Δx(-2/3)-Δx(2/3), Δz(-2/3)+Δz(2/3), Δx(-1/3)-Δx(1/3), Δz(-1/3)+Δz(1/3), Δz(0)} which carries 7 $\Gamma_1$,

- the span of {Δy(-1)-Δy(1), Δy(-2/3)-Δy(2/3), Δy(-1/3)-Δy(1/3)} which carries 3 $\Gamma_2$,

- the span of {Δx(-1)+Δx(1), Δz(-1)-Δz(1), Δx(-2/3)+Δx(2/3), Δz(-2/3)-Δz(2/3), Δx(1/3)+Δx(1/3), Δz(-1/3)-Δz(1/3), Δx(0)} which carries 7 $\Gamma_3$, and

- the span of {Δy(-1)+Δy(1), Δy(-2/3)+Δy(2/3), Δy(-1/3)+Δy(1/3), Δy(0)} which carries 4 $\Gamma_4$.

where Δx(L) is the unit displacement of the oxygen atom in the image at λ=L along x (or y or z). These four subspaces correspond to symmetry breaking to *m*m2*, 2*, m, and m* respectively (these are the kernels of each irrep). Just like how an ordinary symmetry-adapted basis for a static structure would put the force constants matrix in block diagonal form, this basis will as well for the generalization "force constants matrix" that includes the nudged elastic band forces on the path; this would have blocks of 7, 3, 7, and 4 rows corresponding to each of the four subspace specified above. For this particular path, the first three blocks should be positive definite (stable, similar to having positive squared frequency with phonons of static structures). The final block, corresponding to the $\Gamma_4$ irrep, has one or more negative eigenvalues, indicating instability.

**Supplementary Discussion 2: Additional analysis of Figure 3: The consequences of distortion symmetry and balanced forces for NEB calculations**

The initial path created by linearly interpolating between $\lambda = -1$ and $\lambda = +1$ has $m^*m2^*$, or more specifically $m_x^*m_y2_z^*$ symmetry where the subscripts represent axis associated with the operation: $m_x^*$ is a starred mirror whose normal is along x, $m_y$ is a mirror whose normal is along y, and $2_z^*$ is a starred two-fold axis along z (see compass on lower left in Supplementary Figure 2). Applying Neumann's principle to the force on the oxygen atom gives the following results:

- $m_y$ $F_y(\lambda) = -F_y(\lambda) = F_y(\lambda)$, hence $F_y(\lambda) = 0$ (see **red** in Supplementary Fig. 2)
- $m_x^*$ $F_x(\lambda) = -F_x(-\lambda) = F_x(\lambda)$, hence $F_x(\lambda)$ is an odd function of $\lambda$ (see **blue**)
- $2_z^*$ $F_z(\lambda) = F_z(-\lambda) = F_z(\lambda)$, hence $F_z(\lambda)$ is an even function of $\lambda$ (see **green**)

Clearly, the forces on both the initial and converged path are consistent with these in Supplementary Figure 2. The forces of one iteration are used to update the positions for the next iteration, thus guaranteeing that the $m_x^*m_y2_z^*$ symmetry cannot be broken in any subsequent iteration.

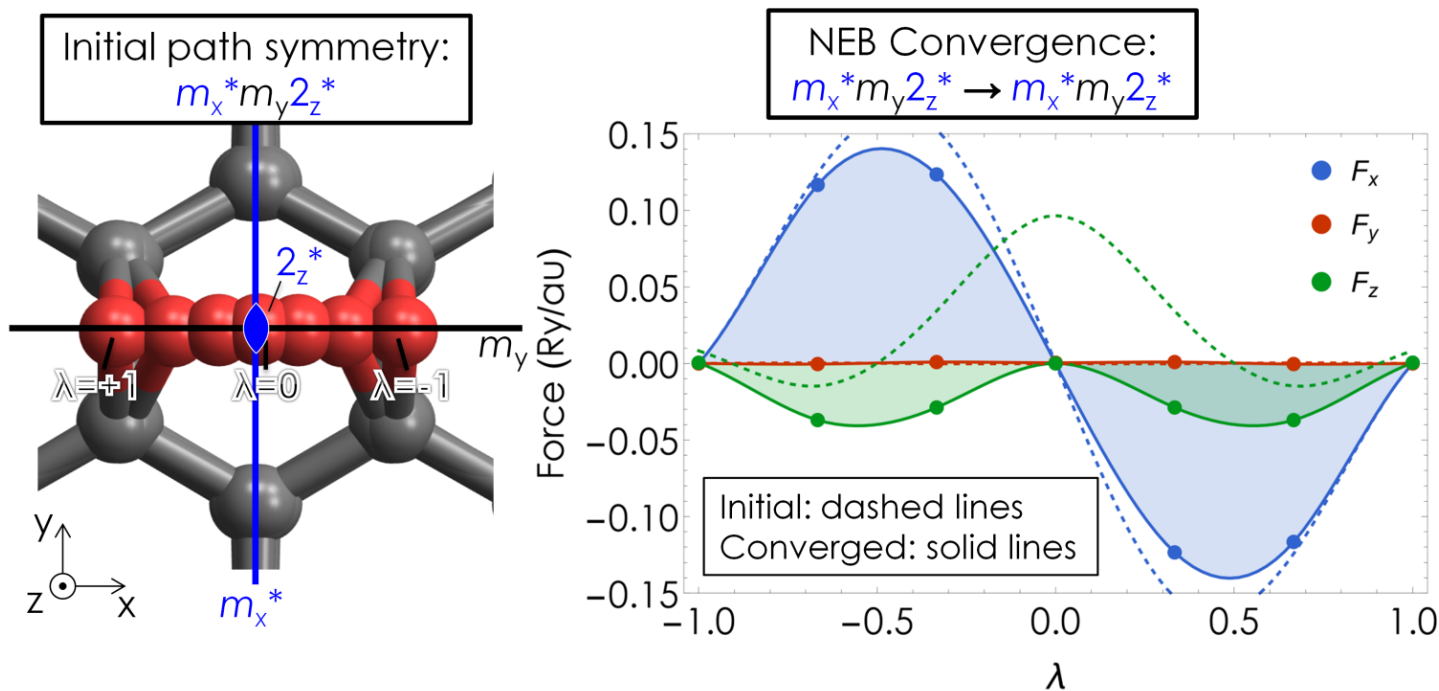

**Supplementary Figure 2: Electrostatic forces on oxygen for the $m^*m2^*$ path.** Left: superimposed images along path after NEB convergence with symmetry diagram overlaid.

Right: x, y, and z component of the force on oxygen for the initial path (dashed) and the final converged path (solid with axis filling).

In Supplementary Figure 3, we deliberately break the $m_x^*m_y2_z^*$ symmetry with a sinusoidal perturbation (the green curve is exaggerated; the maximum displacement was 0.1 Angstrom). Note that the forces on the initial path are similar to the unperturbed case, because the perturbation is small, but slightly break the previous symmetry. The perturbation is such that the path retains $2_z^*$ symmetry. Again applying Neumann's principle to the force on the oxygen atom gives the following results:

- $2_z^*$ $F_y(\lambda) = -F_y(-\lambda) = F_y(\lambda)$, hence $F_y(\lambda)$ is an odd function of $\lambda$ (see **red**)
- $2_z^*$ $F_x(\lambda) = -F_x(-\lambda) = F_x(\lambda)$, hence $F_x(\lambda)$ is an odd function of $\lambda$ (see **blue**)
- $2_z^*$ $F_z(\lambda) = F_z(-\lambda) = F_z(\lambda)$, hence $F_z(\lambda)$ is an even function of $\lambda$ (see **green**)

Again, this is consistent with the forces on both the initial and converged path in Supplementary Fig. 3. The $2_z^*$ symmetry in the initial guess prevents NEB iterations from finding a significantly lower transition state (TS).

**Supplementary Figure 3: Electrostatic forces on oxygen for the 2* perturbed path.** Left: perturbation from m*m2* path (green curve) with 2* symmetry diagram overlaid. Right: x, y, and z component of the force on oxygen for the initial path (dashed) and the final converged path (solid with axis filling).

In Supplementary Figure 4, we deliberately break the $m_x^*m_y2_z^*$ symmetry to trivial symmetry (the green curve is exaggerated; the maximum displacement was 0.18 Angstrom). Note that the forces on the initial path are similar to the unperturbed case, because the perturbation is small, but slightly break all previous symmetry. NEB iterations drive the path to the much lower energy m* path seen in Figure 3d of the main text.

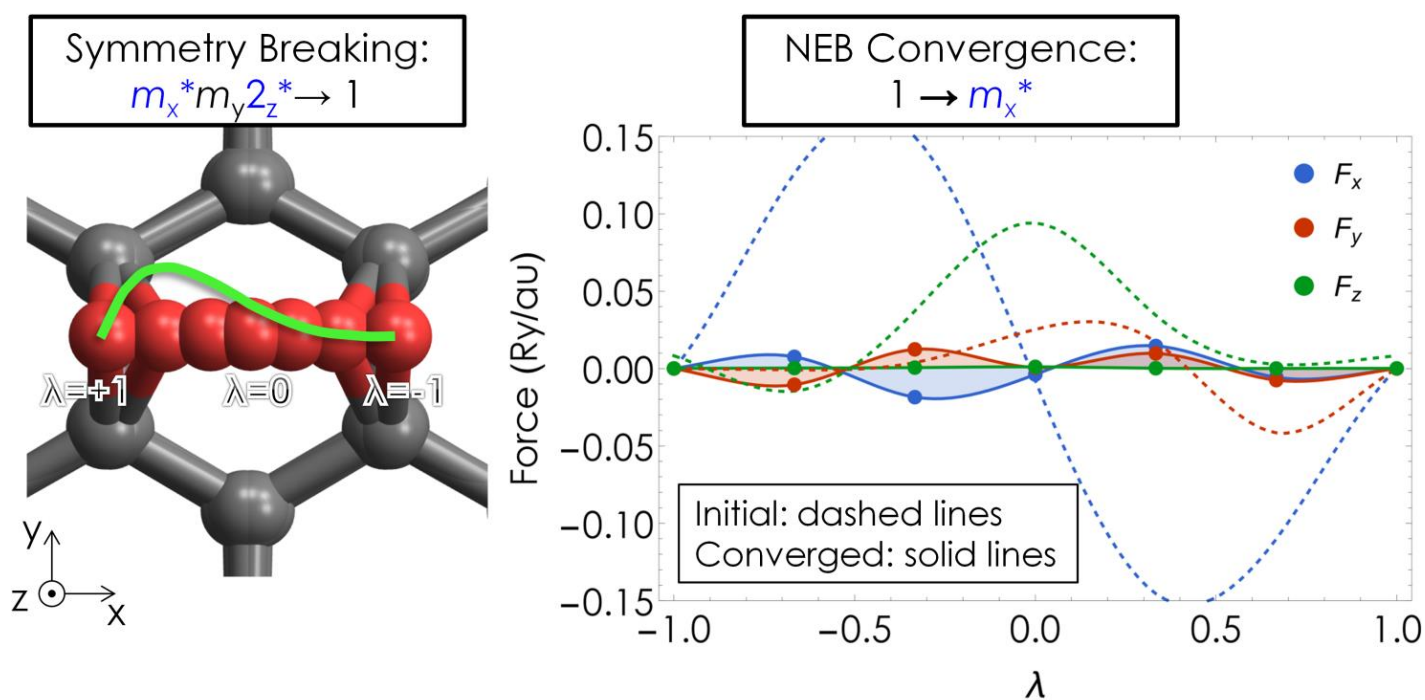

**Supplementary Figure 4: Electrostatic forces on oxygen for the trivial symmetry (1) perturbed path.** Left: perturbation from m*m2* path (green curve). Right: x, y, and z component of the force on oxygen for the initial path (dashed) and the final converged path (solid with axis filling).

Supplementary Figure 5 shows two other ways of breaking *m*m2** symmetry. We expect that NEB would drive the *m** initial path (left side of Supplementary Fig. 5) to the same low energy *m** path seen in Figure 3d of the main text. The *m* initial path (right side of Supplementary Fig. 5) should not be able to converge to the low energy *m** path because *m* is not a subgroup of *m** and NEB iterations must conserve distortion symmetry.

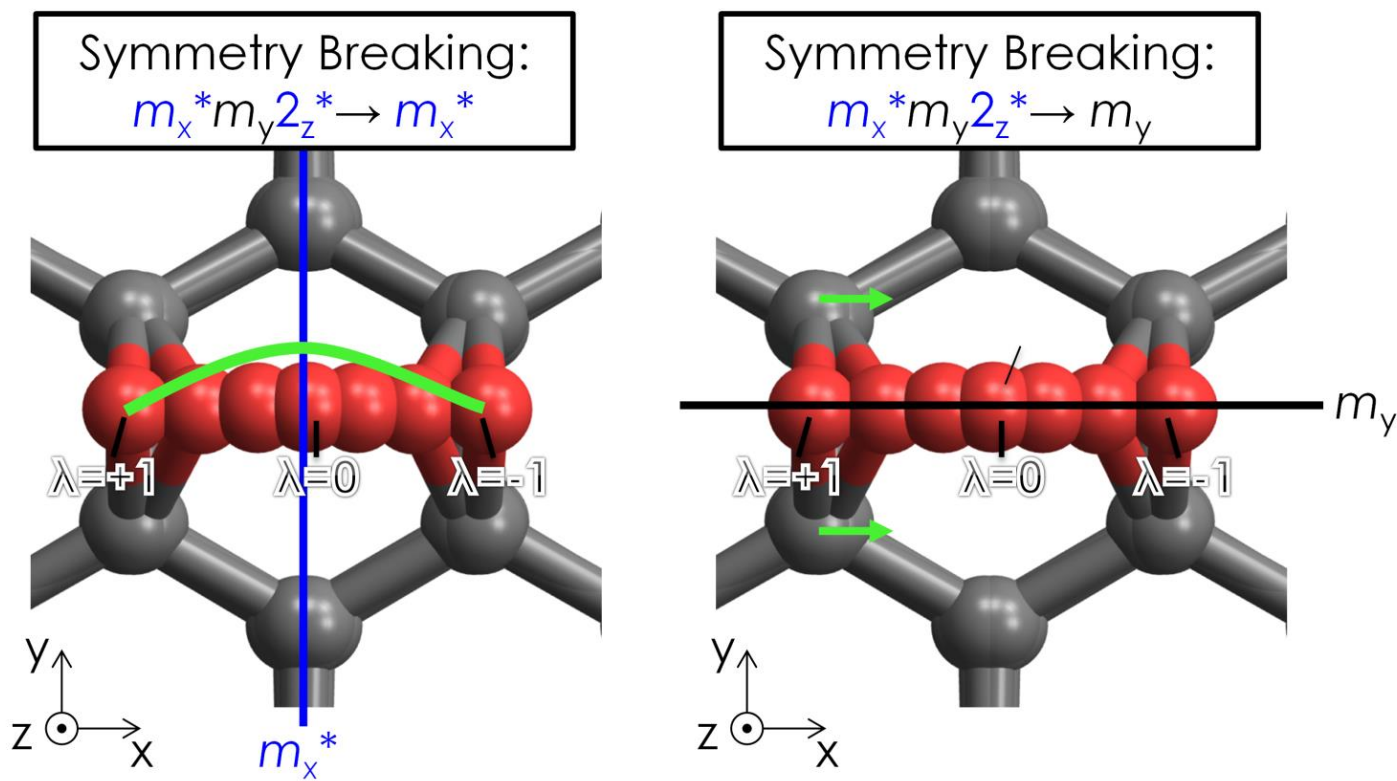

**Supplementary Figure 5: Possible *m*** **and *m* perturbations.** Left: perturbation from *m*m2** path (green curve) to an *m** symmetry path. Right: perturbation from *m*m2** path (green arrows indicating carbon displacements) to an *m* symmetry path.

**Supplementary Discussion 3: Example of 1\* switching handedness for a distortion of quartz**

a

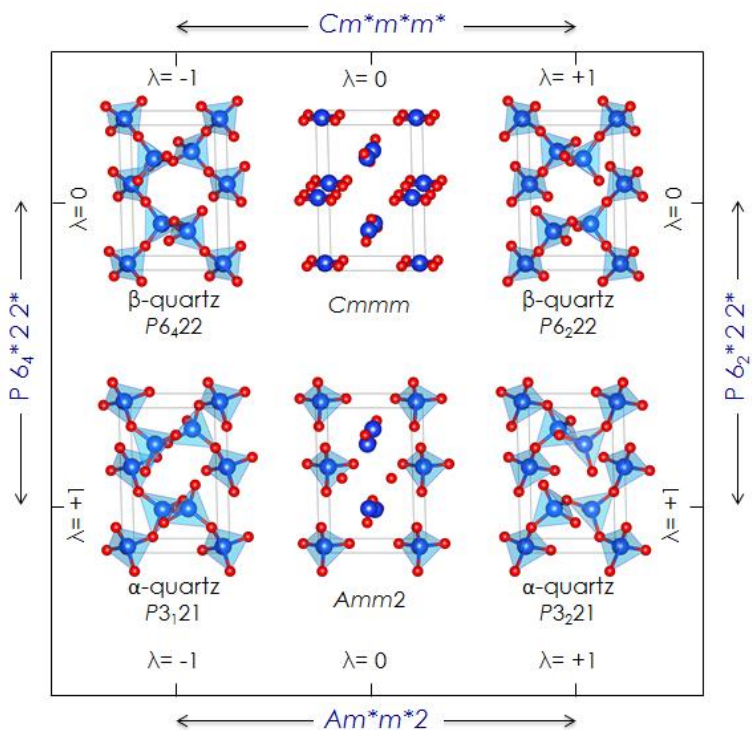

b

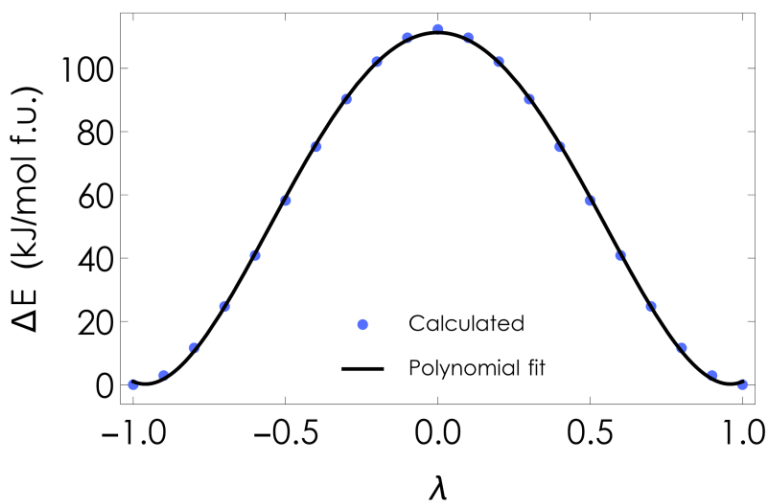

**Supplementary Figure 6:** **a** Some distortion pathways and their distortion symmetry between left and right handed variants of α-quartz and β-quartz. **b** The energy of the $P6_4^*22^*$ distortion as a function of $\lambda$.

Supplementary Figure 6 shows the example of quartz, a common crystal that is found in left- and right- handed configurations, and is commonly used in watches and clocks as a crystal oscillator by using its piezoelectric effect. Trigonal α-quartz transforms into hexagonal β-quartz at 573°C, into hexagonal β-tridymite at 870°C and to cubic β-cristobalite at 1470°C. If β-quartz is considered a parent, the distortion to α-quartz has $P6_2$*22* symmetry for left-handed quartz and $P6_4$*22* symmetry for right-handed quartz (this is the distortion depicted in Fig. 5a of the main text). Note that $\lambda = +1$ and $\lambda = -1$ are different variants of α-quartz with β-quartz structurally intermediate between the two. In this instance, 1* does not reverse the handedness of the structures. However, if an appropriate parent is chosen, one can also transform between left- and right-handed α-quartz, as well as between left- and right- handed β-quartz. For α-quartz, our path has A*m*m*2 symmetry. For β-quartz, our path has C*m*m*m* symmetry. This demonstrates a potentially surprising property of 1*, namely, that it can, for carefully selected paths, 1* reverse the handedness of a crystal. However, our $\lambda = 0$ structure is clearly not physically reasonably for either the A*m*m*2 symmetry path or the C*m*m*m* symmetry path so it is very unlikely that it could be activated experimentally in practice. Other choices of paths are possible with the same or different distortion symmetries so there may be other paths with more reasonable transition states.

More generally, one can find distortion groups describing transformation between any two enantiomorphic structures (related by mirror) by choosing an appropriate parent that is intermediate between the two. Multiple such parents are possible, in principle. These ideas are also applicable to liquid crystals which can switch between left- and right- handed enantiomorphs under an electric field, a property that is utilized in computer displays. Supplementary Figure 6b shows that, as with the examples given in the main text, the energy of the $P6_4$*22* distortion is symmetric with respect to λ due to the starred symmetry.

**Supplementary Discussion 4: Simple NEB example to demonstrate the effect of distortion symmetry on NEB convergence**

In the provided Mathematica Notebook file, "Simple_NEB.nb", an example 2D potential energy surface (PES) is described and a simple implementation of the nudged elastic band (NEB) method is included. This PES is given as:

$$-4.07144 + 0.2e^{-x^2-4y^2} + \text{Cos}[x]\text{Cos}[y] + \text{Cosh}[\frac{x}{2}] + 3\text{Cosh}[\frac{y}{2}]$$

"Simple_NEB.nb" contains dynamic and interactive plots that show what happens to the initial guess path as the NEB method iterates. Supplementary Figure 7 shows an example of what it might look like starting from a path with trivial symmetry after 25 iterations.

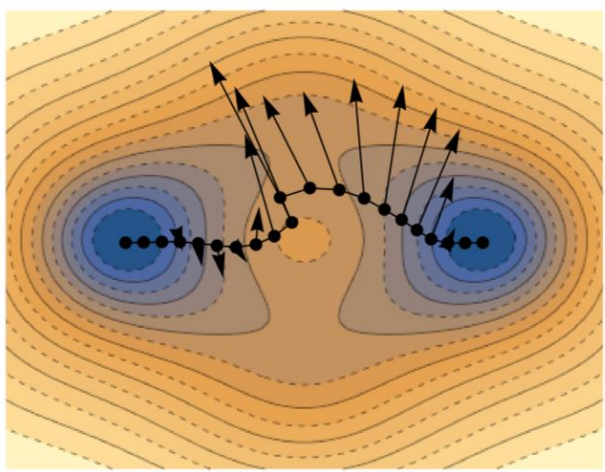
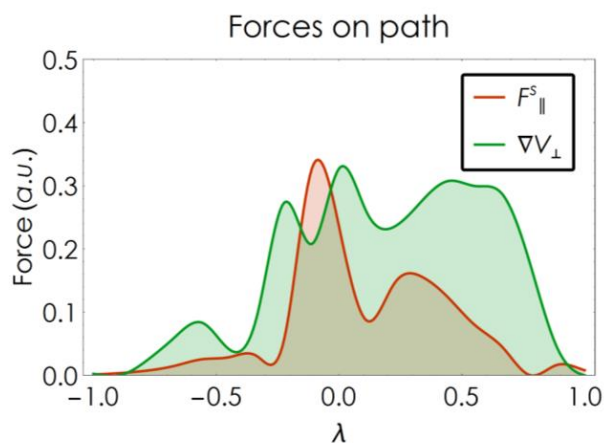
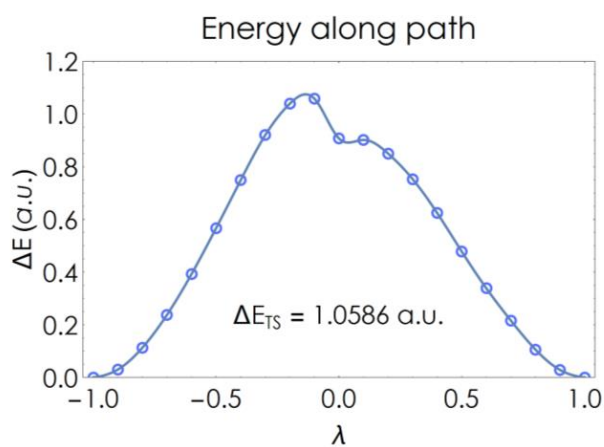
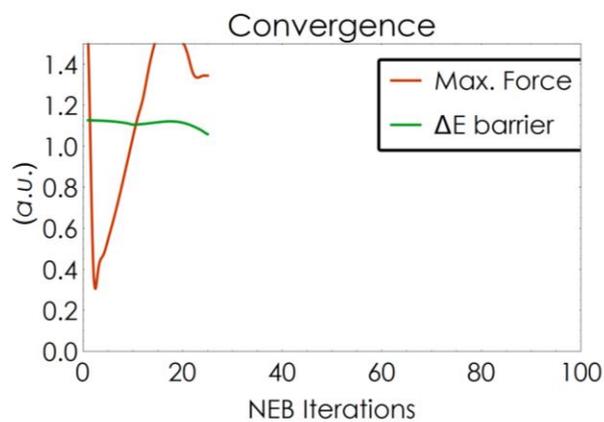

**Supplementary Figure 7:** Top left is the path on the PES with force vectors plotted as arrows, top right is the magnitude of the two components of the forces on the path (elastic band force is red, gradient force is green), lower left is the current energy profile of the path and transition state energy, and lower right is the convergence of the forces and the energy barrier as a function of iterations.

Using this implementation and PES, we tested the idea that applying distortion symmetry should result in more rapid convergence. Our implementation was based on the explaination of the Nudged Elastic Band method given by Jónsson et al. [3]. Starting from a straight path, we generated 100,000 initial paths with trivial symmetry and 100,000 initial paths with *m\** symmetry, which are the conventional symmetry and distortion symmetry of the MEPs in this example. The details of how these were randomly generated are in the "Simple_NEB.nb" file. For each initial path, we ran our NEB implementation until the forces fell below a chosen convergence threshold. The results are summarized in Supplementary Figure 8. Note that convergence is considerably more rapid with distortion symmetry in this example. The conventional symmetry paths typically took more than twice as long; the average number of iterations was about 443.8 for conventional symmetry and 190.8 for distortion symmetry. Symmetrizing using the correct distortion symmetry reduced the number of NEB iterations needed in 98.97% of our test cases and by a factor of 2.3 on average. Also note that if a straight path was provided as the initial guess, convergence to a MEP would not be possible in this example. Consequently, at least for this example, understanding distortion symmetry is crucial for achieving good results.

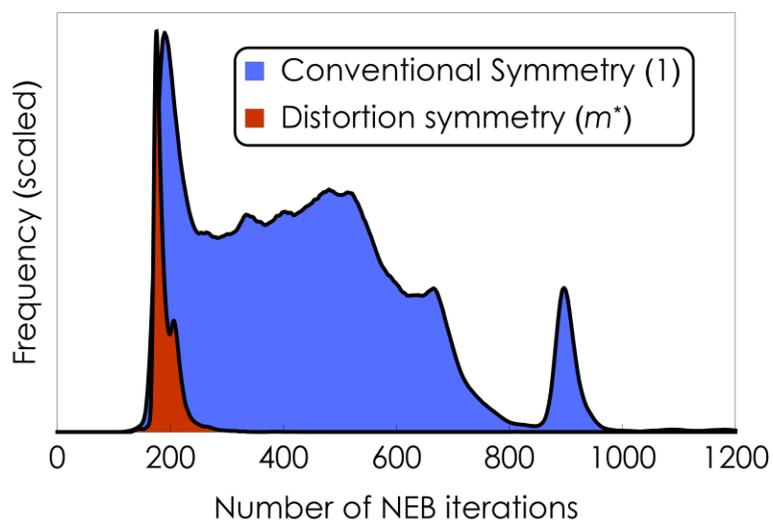

**Supplementary Figure 8:** Smoothed and rescaled histograms of showing the frequency of NEB convergence at a given number of iterations in this example system for initial paths with *m\** or trivial symmetry (1). The two curves are rescaled to the same maximum height.

## Supplementary Discussion 5: The application of distortion symmetry to Berry phase calculations

Davide Ceresoli and Erio Tosatti[4] give the Berry phase along a path as:

$$\gamma \cong -\text{Im} \log \det \prod_{\xi=0}^{N-1} \langle \psi_\xi | \psi_{\xi+1} \rangle$$

$\xi = 0$ through $\xi = N$ are the indices of discrete images along the path (i.e. each corresponds to a set of nuclear positions which can be used to compute the ground state electronic structure, $\psi_\xi$). Ceresoli & Tosatti use $\gamma$ to compute "magnetic screening" (which is $\gamma$ divided by another factor) for some rotations and pseudorotations of simple molecules. The path they describe as "the pseudorotation of benzene" is of particular interest to us. Ceresoli & Tosatti describe this path as having "small magnetic screening" due to the "nearly exact balance of positive and negative currents". We propose that the balance is *perfectly exact* due to starred symmetry. This pseudorotation of benzene has *m\*m\*m* symmetry and we claim that paths with *any* starred symmetries have zero Berry phase (excluding pathological cases like discontinuous paths).

Consider that the product expressed in the expression for Berry phase expands like so:

$$\prod_{\xi=0}^{N-1} \langle \psi_\xi | \psi_{\xi+1} \rangle = \langle \psi_0 | \psi_1 \rangle \langle \psi_1 | \psi_2 \rangle \dots \langle \psi_{N-2} | \psi_{N-1} \rangle \langle \psi_{N-1} | \psi_N \rangle$$

1* reverses the path such that the last image (at $\xi = N$) becomes the first image (at $\xi = 0$) and vice versa. Thus:

$$1^*(\langle \psi_0 | \psi_1 \rangle \langle \psi_1 | \psi_2 \rangle \dots \langle \psi_{N-2} | \psi_{N-1} \rangle \langle \psi_{N-1} | \psi_N \rangle) = \langle \psi_N | \psi_{N-1} \rangle \langle \psi_{N-1} | \psi_{N-2} \rangle \dots \langle \psi_2 | \psi_1 \rangle \langle \psi_1 | \psi_0 \rangle$$

Putting this into the formula for the Berry phase and simplifying leads to the conclusion that $1^*\gamma = -\gamma$ for $|\gamma| < \pi$. Thus, for a path with $1^*$ symmetry, $\gamma = 0$ (assuming $|\gamma| < \pi$). Because $\gamma$ is invariant under rotations, mirrors, and translations (generally, Euclidean motions), any starred symmetry gives the same result, including the starred mirrors of *m\*m\*m* symmetry group of the benzene pseudorotation.

**Supplementary Table 1: Examples of published studies that could have benefited from the application of distortion groups.**

| Reference | Figure(s) with starred symmetry consequences | Description and notes |
|---|---|---|
| Durinck, J., Legris, A. & Cordier, P. Pressure sensitivity of olivine slip systems: First-principle calculations of generalised stacking faults. Phys. Chem. Miner. 32, 646–654 (2005). | 3,4,5,6 | Stacking faults in olivine. All figures that should be symmetric due to starred symmetry show some degree of asymmetric error, Fig. 6 in particular. In part, this is due to having an even number of image. |
| Dai, Y., Ni, S., Li, Z. & Yang, J. Diffusion and desorption of oxygen atoms on graphene. J. physics. Condens. matter. 25, 405301 (2013). | 3 | Both Fig. 3a and 3b should be symmetric due to starred symmetry. Fig. 3a is a particularly extreme example of asymmetry error. The path taken in Fig. 3b is unstable and is balanced by symmetry as discussed in the main text. |
| Zhou, B., Shi, H., Cao, R., Zhang, X. & Jiang, Z. Theoretical study on the initial stage of Magnesium battery based on V2O5 cathode. Phys. Chem. Chem. Phys. 16, 18578–18585 (2014). | 3 | Diffusion of Li and Mg in $V_2O_5$. The energy profiles show significant asymmetry error. |
| Ma, W. Y., Zhou, B., Wang, J. F., Zhang, X. D. & Jiang, Z. Y. Effect of oxygen vacancy on Li-ion diffusion in a V2O5 cathode: a first-principles study. J. Phys. D. Appl. Phys. 46, 105306 (2013). | 4a,c | Calculated energy for three different diffusion pathways for Li in $V_2O_5$ using NEB. Paths A and C have starred symmetry and thus figures 4a and 4c are symmetric. Path B does not have starred symmetry and thus figure 4b is asymmetric. |
| Morgan, B. J. & Watson, G. W. GGA+U description of lithium intercalation into anatase TiO2. Phys. Rev. B - Condens. Matter Mater. Phys. 82, 1–11 (2010). | 8b,c | Calculated energy of Li diffusion pathway in $TiO_2$ using NEB. The pathways for Fig 8b and 8c have starred symmetry and thus the energy plot is symmetric. Fig 9 depicts this pathway. |

| Reference | Figure | Description |
|---|---|---|
| Yang, J., Hu, W. & Tang, J. Surface self-diffusion behavior of individual tungsten adatoms on rhombohedral clusters. J. Phys. Condens. Matter 23, 395004 (2011). | 3 | NEB study of diffusion of tungsten adatom on tungsten cluster surface. Figure 3 shows symmetric energy due to starred symmetry. |
| Tsuru, T. et al. Solution softening in magnesium alloys: the effect of solid solutions on the dislocation core structure and nonbasal slip. J. Phys. Condens. Matter 25, 022202 (2013). | 1c and 3a-d | Energy versus slip in magnesium alloys. Energy is symmetric for prismatic slip in Figure 1c due to starred symmetry. Symmetric energy in Figure 3a and 3b due to starred symmetry. The restoring force is antisymmetric in Figure 3c and 3d due to starred symmetry. |
| Mulroue, J., Uberuaga, B. P. & Duffy, D. M. Charge localization on the hexa-interstitial cluster in MgO. J Phys Condens Matter 25, 65502 (2013). | 6 | Diffusion path of MgO/$O_3$ cluster in MgO. Figure 6 is symmetric due to starred symmetry, but has small asymmetric errors. |
| Gröger, R. & Vitek, V. Constrained nudged elastic band calculation of the Peierls barrier with atomic relaxations. Model. Simul. Mater. Sci. Eng. 20, 035019 (2012). | 2,3,4, and 5b | Constrained nudged elastic band calculation of the Peierls barrier with atomic relaxations. There are small but apparent deviations from symmetry (numerical errors). Fig. 5b shows only half of the pathway, due to symmetry. |
| Schusteritsch, G. & Kaxiras, E. Sulfur-induced embrittlement of nickel: a first-principles study. Model. Simul. Mater. Sci. Eng. 20, 065007 (2012). | 7 | Stacking faults in Ni. Figure 7b (Theta=0.0) and 7c(Theta=0.0) are antisymmetric due to symmetry. |
| Wu, M. H., Liu, X. H., Gu, J. F. & Jin, Z. H. DFT study of nitrogen – vacancy complexions in ( fcc ) Fe. Model. Simul. Mater. Sci. Eng. 22, 055004 (2014). | 4a | N diffusion in fcc Fe. Fig. 4a is symmetric due to starred symmetry. |

| Reference | Figure | Description |
|---|---|---|
| Lawrence, B., Sinclair, C. W. & Perez, M. Carbon diffusion in supersaturated ferrite: a comparison of mean-field and atomistic predictions. Model. Simul. Mater. Sci. Eng. 22, 065003 (2014). | 3 | Carbon diffusion in supersaturated ferrite. Figure 3 ($x_c$ = 0) is symmetric due to starred symmetry. |
| Samanta, A. & Weinan, E. Optimization-based string method for finding minimum energy path. Commun. Comput. Phys. 14, 265–275 (2013). | 2,3 | Ad-atom diffusion on copper surface. Figs. 2 and 3 are symmetric due to starred symmetry. |
| Zelený, M. et al. Ab initio study of Cu diffusion in α-cristobalite. New J. Phys. 14, (2012). | N/A | Cu diffusion in cristobalite. None of these figures are symmetric due to starred symmetry. Some are nearly symmetric barriers, but not exact. This is an example that shows that apparent symmetry in energy plots is not necessarily due to distortion symmetry (and therefore not exact). |
| Mulroue, J. & Duffy, D. An ab initio study of the effect of charge localization on oxygen defect formation and migration energies in magnesium oxide. Proc. R. Soc. London. A. Math. Phys. Sci. 467, 2054–2065 (2011). | 2,3 | An ab initio study of the effect of charge localization on oxygen defect formation and migration energies in magnesium oxide. Figs. 2 and 3 are symmetric due to starred symmetry. |
| Śpiewak, P. & Kurzydłowski, K. J. Formation and migration energies of the vacancy in Si calculated using the HSE06 range-separated hybrid functional. Phys. Rev. B - Condens. Matter Mater. Phys. 88, 1–6 (2013). | 3 | Vacancy diffusion pathway, CI-NEB. Fig. 3 is symmetric due to starred symmetry. |
| Kumagai, T. et al. H-atom relay reactions in real space. Nat. Mater. 11, 167–172 (2011). | 4 | H-atom relay reactions in real space. Fig. 4 is symmetric due to starred symmetry. |

| Reference | Figure | Description |
|---|---|---|
| Pizzagalli, L. et al. A new parametrization of the Stillinger-Weber potential for an improved description of defects and plasticity of silicon. J. Phys. Condens. Matter 25, 055801 (2013). | 2 | Stacking fault energy. Figure 2 is symmetric due to starred symmetry. |
| Pei, Z. et al. Ab initio and atomistic study of generalized stacking fault energies in Mg and Mg-Y alloys. New J. Phys. 15, 0–19 (2013). | 3,4,5 | Stacking fault energy in Mg and Mg-Y alloys. Figs. 3, 4, and 5 are symmetric due to starred symmetry. |
| Ho, G., Ong, M. T., Caspersen, K. J. & Carter, E. a. Energetics and kinetics of vacancy diffusion and aggregation in shocked aluminium via orbital-free density functional theory. Phys. Chem. Chem. Phys. 9, 4951–4966 (2007). | 4 | Diffusion in aluminum. Figure 4 is symmetric due to starred symmetry. |
| Panduwinata, D. & Gale, J. D. A first principles investigation of lithium intercalation in TiO2-B. J. Mater. Chem. 19, 3931 (2009). | N/A | Lithium intercalation in TiO2-B. Shows the result of reversed pathways, e.g. Fig. 5. |
| Berry, S. Correlation of rates of intramolecular tunneling processes, with application to some group V compounds. J. o Chem. Phys. 32, 933–938 (1960). | 2 | Arrows depicting $PF_5$ pseudorotation in Fig. 2. As noted in the main text, these arrows are related by symmetry. |
| Kushima, A. et al. Interstitialcy diffusion of oxygen in tetragonal La2CoO(4+δ). Phys. Chem. Chem. Phys. 13, 2242–2249 (2011). | 2a | Diffusion of oxygen in $La_2CoO_4$. Figure 2b shows Initial to Saddle images but not Saddle to Final, implying the intuitive application of symmetry. |

| Reference | Figure | Notes |
|---|---|---|
| Li, W. et al. Li+ ion conductivity and diffusion mechanism in α-Li3N and β-Li3N. Energy Environ. Sci. 3, 1524 (2010). | 7 | Diffusion in $Li_3N$. Pathways depicted in Fig. 6. Figure 7 is symmetric due to starred symmetry. |
| Yildirim, H., Greeley, J. P. & Sankaranarayanan, S. K. R. S. The effect of concentration on Li diffusivity and conductivity in rutile TiO2. Phys. Chem. Chem. Phys. 14, 4565 (2012). | 3 | Li diffusion in $TiO_2$. Figure 3 is symmetric due to starred symmetry. |
| Ye, X.-J., Liu, C.-S., Jia, R., Zeng, Z. & Zhong, W. How does the boron concentration affect hydrogen storage in lithium decorated zero- and two-dimensional boron-carbon compounds? Phys. Chem. Chem. Phys. 15, 2507–13 (2013). | 5 | Li diffusion on $B_8C_{24}$ and $B_{24}C_{12}$. The path on $B_8C_{24}$ has starred symmetry and thus the energy plot in Fig. 5c should be symmetric but is not, presumably due to errors. The path on $B_{24}C_{12}$ does not have starred symmetry and the asymmetry of Fig. 5d is consistent with this. |
| Han, J. W. & Yildiz, B. Mechanism for enhanced oxygen reduction kinetics at the (La,Sr)CoO3−δ/(La,Sr)2CoO4+δ hetero-interface. Energy Environ. Sci. 5, 8598 (2012). | 3e | O diffusion in $La_2CoO_4$. Figure 3e is symmetric due to starred symmetry. |
| Kuhlman, T. S., Glover, W. J., Mori, T., Møller, K. B. & Martínez, T. J. Between ethylene and polyenes - the non-adiabatic dynamics of cis-dienes. Faraday Discuss. 157, 193 (2012). | 5 | NEB, molecular transitions. Figure 5 is symmetric due to starred symmetry. |
| Matsunaga, K. & Toyoura, K. First-principles analysis of oxide-ion conduction mechanism in lanthanum silicate. J. Mater. Chem. 22, 7265 (2012). | N/A | Oxygen diffusion in lanthanum silicate. Very interesting pathway from O5-0 to symmetry equivalent O5-0 site depicted in Fig. 9 and Fig. 10. Shows how equivalence of initial and final states (as noted in the captions) does not guarantee symmetry. This can be seen from the labels given along the O5-0 to O5-0 pathways, e.g. in Fig. 10 the pathway goes from O5-0 to O5-s2 to O5-s1 to O5-0. Because O5-s2 and O5-s1 are inequivalent, it is impossible to superimpose this pathway with its reverse and therefore there is no starred symmetry. |
| Islam, M. S. & Fisher, C. a J. Lithium and sodium battery cathode materials: computational insights into voltage, diffusion and nanostructural properties. Chem. Soc. Rev. 43, 185–204 (2014). | 7 | Migration pathway for Li in $LiFePO_4$. Fig.5 also depicts two different kinds of hops that occur in $Li_xCoO_2$, we note that the first (Fig. 5a) has starred symmetry but the other does not. |
| Cai, Y. et al. Constructing metallic nanoroads on a $MoS_2$ monolayer via hydrogenation. Nanoscale 6, 1691–7 (2014). | 2 | Diffusion of hydrogen atom on $MoS_2$. Figure 2 is symmetric due to starred symmetry. |
| Murugesan, S. et al. Wide electrochemical window ionic salt for use in electropositive metal electrodeposition and solid state Li-ion batteries. J. Mater. Chem. A 2, 2194 (2014). | N/A | Fig. 6 shows another example of a minimum energy pathway that is not superimposable with its reverse and therefore does not contain starred symmetry. |
| Gao, Y. et al. Improved electron/Li-ion transport and oxygen stability of Mo-doped Li2MnO3. J. Mater. Chem. A 2, 4811 (2014). | 5a,b | Various migration pathways in $Li_2MnO_3$. Path 3 and 4 have starred symmetry, the rest do not. |

| Reference | Figures | Notes |
|---|---|---|
| Wu, J. et al. Tavorite-FeSO4F as a potential cathode material for Mg ion batteries: a first principles calculation. Phys. Chem. Chem. Phys. 16, 22974–22978 (2014). | 3,4 | Mg diffusion in $Mg_{0.5}FeSO_4F$. L1 and L2 pathways (Fig. 3) both have starred symmetry. Small deviations from this symmetry are apparent in Fig.4, e.g. the 5th and 13th points of Fig. 4a should be at the same height. Presumably this is due to errors in the calculations. |
| Su, J., Pei, Y. & Wang, X. Ab initio study of graphene-like monolayer molybdenum disulfide as a promising anode material for rechargeable sodium ion batteries. RSC Adv. 4, 43183–43188 (2014). | 4 | Na migration path on MoS2. Very clear example with superimposed images showing the pathway. Both pathways in Figure 4 have starred symmetry. |
| Ling, C., Zhang, R. & Mizuno, F. Phase stability and its impact on the electrochemical performance of VOPO4 and LiVOPO4. J. Mater. Chem. A 2, 12330 (2014). | 7 | Li diffusion in $LiVOPO_4$. Figure 7 is symmetric due to starred symmetry. |
| Mo, Y., Ong, S. P. & Ceder, G. Insights into Diffusion Mechanisms in P2 Layered Oxide Materials by First-Principles Calculations. Chem. Mater. 26, 5208–5214 (2014). | 3 | Na diffusion in $NaCoO_2$. Figure 3 is symmetric due to starred symmetry. |
| Nishimura, S. et al. Experimental visualization of lithium diffusion in LixFePO4. Nat. Mater. 7, 707–711 (2008). | 1c,d and 3 | Figure 3 shows the expected diffusion path of Li in $LiFePO_4$. Depicted as continuous motion. Figure 1c,d show two different pathways. Both have clear starred symmetry. Considering only the finite structures depicted, 1c has m*m2* symmetry and 1d has 2*/m symmetry. |

| Reference | Figures | Notes |
|---|---|---|
| Milas, I., Hinnemann, B. & Carter, E. a. Diffusion of Al, O, Pt, Hf, and Y atoms on α-Al2O3(0001): implications for the role of alloying elements in thermal barrier coatings. J. Mater. Chem. 21, 1447 (2011). | 3a,d,e | Diffusion on $Al_2O_3$ surface. Fig. 3a,d, and e are symmetric due to starred symmetry. |
| Marinica, M.-C. et al. Interatomic potentials for modelling radiation defects and dislocations in tungsten. J. Phys. Condens. Matter 25, 395502 (2013). | 2,4,6 | Figs. 2, 4, and 6 are symmetric due to starred symmetry. |
| Chang, H. et al. Single adatom dynamics at monatomic steps of free-standing few-layer reduced graphene. Sci. Rep. 4, 6037 (2014). | 4b | Pathways in Fig. 4a and c have equivalent endpoints, but do not have starred symmetry (the approximate symmetry of the energy in Fig. 4c is not due to starred symmetry). The pathway in Fig. 4b should have starred symmetry but there are clear deviations in the energy (note 2nd and 2nd to last points). The starred symmetry suggests this is an error, maybe due to using an even number of images. |
| Wang, C. et al. Single Adatom Adsorption and Diffusion on Fe Surfaces. Journal of Modern Physics 02, 1067–1072 (2011). | 2,3,4 | Adatom Diffusion on Fe surfaces. Small asymmetry in Fig 2 due to choosing an even number of points. |

| Reference | | Description |
|---|---|---|
| Schegoleva, L. N. & Beregovaya, I. V. Manifestation of Complicated Structure of Potential Energy Surfaces in Spectral and Chemical Properties of Haloarene Radical Ions. Fluor. Notes 2, (2013). | 10 | Pseudo-rotation of 1,2,3-$F_3C_6H_3^-$. In Figure 10b, starred symmetry has the consequence of making $F_2$ symmetric and $F_1$ and $F_3$ mirror images. This is similar to the $PF_5$ pseudorotation and bond lengths example given in the main text of our work. |
| Park, Y.-U. et al. Tailoring a fluorophosphate as a novel 4 V cathode for lithium-ion batteries. Sci. Rep. 2, 1–7 (2012). | 5 | Diffusion of Na and Li ions in $Na_{1.5}VPO_5F_{0.5}$ and $LiNa_{0.5}VPO_5F_{0.5}$. Fig. 5b and 5d give very clear depictions of the pathways. Fig. 5c and 5e are symmetric due to starred symmetry. |
| Wang, J., Ewing, R. C. & Becker, U. Average structure and local configuration of excess oxygen in UO2+x. Sci. Rep. 4, 4216 (2014). | 1 | Fig. 1 is symmetric due to starred symmetry. |
| Tang, Z. K., Zhang, Y. N., Zhang, D. Y., Lau, W. M. & Liu, L. M. The stability and electronic properties of novel three-dimensional graphene-MoS2 hybrid structure. Sci. Rep. 1–7 (2014). | 5 | Fig. 5 is symmetric due to starred symmetry. |
| Tingaud, D., Nardou, F. & Besson, R. Diffusion in complex ordered alloys: Atomic-scale investigation of NiAl3. Phys. Rev. B - Condens. Matter Mater. Phys. 81, 1–8 (2010). | 3 | Diffusion in NiAl3. Fig. 3 is a comparison of NEB and constrained atom (CA) methods. Both show small asymmetry errors. Fig. 3 is symmetric due to starred symmetry. |
| Shang, S., Hector Jr., L. G., Wang, Y. & Liu, Z. K. Anomalous energy pathway of vacancy migration and self-diffusion in hcp Ti. Phys. Rev. B 83, (2011). | 1 | Diffusion in Ti. Figure 1 is symmetric due to starred symmetry. |
| Dathar, G. K. P., Sheppard, D., Stevenson, K. J. & Henkelman, G. Calculations of Li-ion diffusion in olivine phosphates. Chem. Mater. 23, 4032–4037 (2011). | 1,2 | Li diffusion in olivine phosphates (FePO4 and LiFePO4). All paths with plotted energies are symmetric due to starred symmetry. |
| Fennie, C. J. & Rabe, K. M. Ferroelectric transition in YMnO3 from first principles. Phys. Rev. B - Condens. Matter Mater. Phys. 72, 1–4 (2005). | 2,4 | The distortions in this paper come from following the normal modes of a parent structure. Figure 4 is antisymmetric because of the starred symmetry. |
| Meng, X. et al. Direct visualization of concerted proton tunnelling in a water nanocluster. Nat. Phys. 11, 235–239 (2015). | 3a | Figure 3a is symmetric due to starred symmetry. |